\documentclass[1p]{elsarticle}
\usepackage{amsfonts}
\usepackage{mathtools}
\usepackage{amssymb}
\usepackage{graphicx}
\usepackage{caption}
\usepackage{epstopdf}
\usepackage{color}
\usepackage{pdflscape}

\journal{Journal of Theoretical Biology}

\biboptions{authoryear}

\begin{document}

\begin{frontmatter}

\title{Modelling cross-reactivity and memory in the cellular adaptive immune response to influenza infection in the host}

\author[mathsunimelb]{Ada W. C. Yan}
\author[mathsunimelb]{Pengxing Cao}
\author[msyork,mi2york]{Jane M. Heffernan}
\author[doherty,pophealthunimelb,murdoch]{Jodie McVernon}
\author[microbiolunimelb,monash]{Kylie M. Quinn}
\author[microbiolunimelb,monash]{Nicole L. La Gruta}
\author[who,federation,microbiolunimelb]{Karen L. Laurie}
\author[mathsunimelb,pophealthunimelb,murdoch]{James M. McCaw\corref{mycorrespondingauthor}}
\cortext[mycorrespondingauthor]{Corresponding author}
\ead{jamesm@unimelb.edu.au}

\address[mathsunimelb]{School of Mathematics and Statistics, University of Melbourne, Parkville, VIC 3010, Australia}
\address[msyork]{Department of Mathematics and Statistics, York University, Toronto, Ontario, Canada M3J 1P3}
\address[mi2york]{Modelling Infection and Immunity Lab, Centre for Disease Modelling, York Institute for Health Research, York University, Toronto, Ontario, Canada M3J 1P3}
\address[doherty]{Doherty Epidemiology, Doherty Institute for Infection and Immunity, University of Melbourne, Parkville, VIC 3010, Australia}
\address[pophealthunimelb]{Centre for Epidemiology and Biostatistics, Melbourne School of Population and Global Health, University of Melbourne, Parkville, VIC 3010, Australia}
\address[murdoch]{Modelling and Simulation, Infection and Immunity Theme, Murdoch Children’s Research Institute, Parkville, VIC 3052, Australia}
\address[microbiolunimelb]{Department of Microbiology and Immunology, Doherty Institute for Infection and Immunity, University of Melbourne, Parkville, VIC 3010, Australia}
\address[monash]{Infection and Immunity Program and Department of Biochemistry and Molecular Biology, Biomedicine Discovery Institute, Monash University, Clayton, Victoria 3800, Australia}
\address[who]{WHO Collaborating Centre for Reference and Research on Influenza, Peter Doherty Institute for Infection and Immunity, Melbourne, VIC 3000,
Australia}
\address[federation]{School of Applied and Biomedical Sciences, Federation University, Churchill, VIC 3842, Australia}

\begin{abstract}
The cellular adaptive immune response plays a key role in resolving influenza infection.  Experiments where individuals are successively infected with different strains within a short timeframe provide insight into the underlying viral dynamics and the role of a cross-reactive immune response in resolving an acute infection.  We construct a mathematical model of within-host influenza viral dynamics including three possible factors which determine the strength of the cross-reactive cellular adaptive immune response: the initial naive T cell number, the avidity of the interaction between T cells and the epitopes presented by infected cells, and the epitope abundance per infected cell.  Our model explains the experimentally observed shortening of a second infection when cross-reactivity is present, and shows that memory in the cellular adaptive immune response is necessary to protect against a second infection.
\end{abstract}


\end{frontmatter}

\section{Introduction}
\label{sec:intro}

The immune response plays an important role in the resolution of primary acute influenza infection and prevention of subsequent infection in an individual.  It can be divided broadly into three parts: the innate immune response, which is fast-acting and non-specific in that it protects the host against a wide range of pathogens~\citep{Kreijtz2011}; the humoral adaptive immune response, which is slower-acting but required to clear infection~\citep{Iwasaki1977}; and the cellular adaptive immune response, which shortens infection and prevents severe disease~\citep{Sridhar2013}.  

The adaptive immune response is antigen-specific. The host has a diverse repertoire of B cells (for the humoral response) and T cells (for the cellular response); these target short peptide fragments, or epitopes, presented by infected cells and professional antigen-presenting cells~\citep{Rammensee1993}.  Only B cells and T cells specific to presented epitopes are stimulated by infection, and the protective effects of the adaptive immune response, such as neutralisation of infectious virus and lysis (killing) of infected cells, are directed at those viral epitopes.   In practice, the immune response to a pathogen is directed at a small number of epitopes, which are termed immunodominant~\citep{Yewdell1999}.  Cross-reactivity in the adaptive immune response occurs when epitopes are shared between virus strains, such that B cells and T cells stimulated by one virus strain can protect against infection with another.

Although antibodies form the first line of defence in protection against influenza infection, immunodominant epitopes targeted by the humoral adaptive immune response appear on the hemagglutinin and neuraminidase proteins, surface proteins which rapidly mutate between strains, such that individuals with previous immunity are no longer immune to the mutated strain~\citep{Brown2009}.  As a result, in the very common occurrences of antigenic drift and antigenic shift which respectively result in epidemics and pandemics, significant evidence is accumulating to suggest that CD8$^+$ T cells play a critical protective role.  This is by virtue of the fact that immunodominant epitopes targeted by the cellular adaptive immune response are typically found in internal viral proteins, which are highly conserved between different influenza A subtypes~\citep{Braciale1977,Kees1984,Yewdell1985,Komadina2016}.  Consequently, cross-reactive memory T cells which remain from previous influenza A infections can be re-activated upon subsequent infection with other influenza A viruses.

In humans, T cells which are cross-reactive between different influenza A subtypes have been detected~\citep{Jameson1999,Boon2004,Greenbaum2009}.  The capacity to induce a strong cellular adaptive immune response \citep{McElhaney2006}, the presence of cross-reactive T cells~\citep{McMichael1983} and previous influenza A infection~\citep{Epstein2006} are all correlated with protection against subsequent infection, as reviewed by \citet{Tscharke2015}.  In the context of the 2009 A(H1N1) influenza pandemic, where individuals had no pre-existing cross-reactive antibodies, disease severity was inversely correlated with the frequency of cross-reactive T cells~\citep{Sridhar2013}.  However, direct causal evidence for protection from infection due to cross-reactive T cells is limited in human studies.  Animal studies provide an opportunity to directly observe the impact of recent previous infections on a subsequent infection, including investigation of the role of the cross-reactive cellular adaptive immune response.  

Mice infected with influenza A viruses (H3N2 and/or H1N1), then exposed to a virulent H7N7 strain cleared the H7N7 infection more quickly than naive mice, and this was shown to be mediated by the recall of virus-specific CD8$^+$ T cell memory \citep{Christensen2000}.  Ferrets recently infected with a seasonal strain of influenza A, then exposed to a pandemic strain, had reduced symptoms and viral loads compared to ferrets without prior exposure~\citep{Bodewes2011}; similar results were found in guinea pigs regarding reduced viral load and transmission~\citep{Steel2010}.  \citet{Laurie2010} further showed that two prior infections were more effective than one at preventing infection and onward transmission.

Natural infection with influenza in humans is typically separated by a period of years; on the other hand, animal experiments can be conducted where infections are separated by days, leading to the observation of viral interference, where very recent/ongoing infection with a first strain impacts the time course of infection with a second strain.  These studies provide an opportunity to observe the effect of the cross-reactive cellular adaptive immune response.  Our recent experimental study \citep{Laurie2015} showed that when ferrets are inoculated with two influenza strains with a 1--14 day interval separating the two exposures, the second infection can be delayed, prevented, or shortened by the presence of the first infection.  The experimental data showed that the effects of this temporary immunity depended not only on the interval between exposures to the two viruses, but also on the strains used; delay and prevention of the second infection were observed whether the two infections were with different types of influenza virus (A and B) or different subtypes of influenza~A, but shortening of the second infection was only observed for infection with different subtypes of influenza~A.  We have previously developed within-host viral dynamics models that demonstrated how the innate immune response could be responsible for a delayed or prevented second infection~\citep{Cao2015}. Those models were then extended by incorporating the cellular adaptive immune response to investigate the relationship between CD8$^+$ T cells and recovery time~\citep{Cao2016}.  However, due to its focus on primary viral infection, the model by~\citet{Cao2016} included neither the formation of memory T cells nor cross-reactivity of memory T cells for serologically distinct influenza strains, leaving the effect of those factors on secondary viral infection unexplored.

Within-host viral dynamics models for the cellular adaptive immune response to multiple infections have been developed for a range of pathogens.  For example, in the case of homologous challenge with lymphocytic choriomeningitis virus, \citet{Chao2004} showed that the peak viral load and the recovery time are reduced for the second infection due to the formation of a large pool of memory T cells after primary infection.  The model included the generation of a T cell repertoire through thymic selection, and allowed for avidity to vary between T cells.  However, the model did not extend to multiple viral strains, and cannot be directly applied to influenza infection due to the lack of an antibody response.  An influenza model by \citet{Zarnitsyna2016} showed that a second infection is shortened by the presence of resident T cells and/or central memory T cells; however, this study assumes complete cross-reactivity between the two exposures.  Other within-host influenza dynamics models of cross-reactive cellular adaptive immunity have focused on the emergence of an immunodominance hierarchy for a primary infection~\citep{Luciani2013}, and how decreased viral load during a second infection changes the immunodominance hierarchy~\citep{Handel2008}, rather than focusing on infection outcomes.  However, these models did not include the innate or humoral adaptive immune response, and did not model the formation of memory T cells.  

In this work, we develop a viral dynamics model for cross-reactive cellular adaptive immune responses induced by multiple infections, including the formation of memory T cells.  We examine three factors for changing the strength of the cross-reactive immune response: 

\begin{enumerate}
\item Changing the precursor frequency (initial number) of epitope-specific CD8$^+$ T cells;
\item Changing the avidity of the interaction between CD8$^+$ T cells and the peptide-MHC complex (pMHC) which presents the epitope;
\item Changing the epitope abundance per infected cell (the number of epitope-specific pMHC on the surface of the cell).
\end{enumerate}

We use the model to explain the observation by \citet{Laurie2015} that a shortened second infection only occurs when infection is with heterosubtypic influenza A strains, and only at inter-exposure intervals of sufficient length for memory in the cellular adaptive immune response to develop.  The results show that under the assumption of negligible cross-reactivity in the humoral adaptive immune response, both cross-reactivity in the cellular adaptive immune response and the formation of a memory CD8$^+$ T cell pool are required to reproduce this shortening of the second infection. We also examine the boosting of the immune response due to multiple infections which induce cross-reactive immune responses.

The model for the cellular adaptive immune response is integrated into a viral dynamics model which also includes the innate and humoral adaptive immune responses.  There are a large number of existing within-host influenza viral dynamics models which well describe the viral load for a single infection, but disagree on the importance of different components of the immune response (as reviewed by \citet{Smith2011a,Beauchemin2011,Dobrovolny2013}), due to identifiability issues which arise when fitting such models to viral load data from a single infection~\citep{Smith2010a,Miao2011,Boianelli2015a}.  Knockout experiments, where some components of the immune response are removed by genetic modification, provide one way to quantify the components of the immune response separately; our recent study~\citep{Cao2016} developed a model which reproduces the viral dynamics of three knockout experiments which removed CD8$^+$ T cell and/or antibody responses~\citep{Kris1988,Iwasaki1977}.  The model in our current study is based on this previous model, but includes the formation of memory CD8$^+$ T cells as well as multiple pools of CD8$^+$ T cells which are used to model cross-reactivity between strains.  Although successive infections within a short timeframe rarely arise in natural infection, experiments on this short timeframe allow us to change the initial conditions of the second infection by changing the inter-exposure interval.  This enables independent assessment of the impact of effector and memory CD8$^+$ T cell subsets which are generated during distinct time periods after infection, as well as distinction of the roles of the early innate immune response and subsequent adaptive immune response.

\section{Materials and Methods}
\label{sec:methods}

\subsection{The model}
\label{sec:model}

We model infection by $Q$ influenza strains and the cellular adaptive immune response to $J$ epitopes across the $Q$ strains.  The dynamics of the multi-strain model (Fig.~\ref{fig:flow}A) are described by a set of ordinary differential equations:

\begin{subequations}
\begin{align}
\frac{dT}{dt} &= gT\left(1-\frac{T+\sum_{q=1}^Q I_q}{T_0}\right) - \sum_{q=1}^Q \beta_q V_q T\label{eq:dTdt} \\
\frac{dF}{dt} &= \sum_{q=1}^Q p_{Fq}I_q - \delta_{F} F \label{eq:dFdt}\\
\frac{dI_q}{dt} &= \beta_q V_qT - \left[\delta_{Iq} + \kappa_{Nq}F + \sum_{j=1}^J (\kappa_{Ejq}E_j + \kappa_{\hat{E}jq}\hat{E}_j)\right] I_q,\qquad q = 1,\hdots,Q \label{eq:dIdt}\\
\frac{dV_q}{dt} &= p_{Vq} I_q - (\delta_{Vq}+ \kappa_{Aq} A_q + \beta_q T)V_q. \label{eq:dVdt}
\end{align} 
\label{eq:ode_multistrain}
\end{subequations}

When target cells ($T$) are infected by virions of strain $q$ ($V_q$), they become infected cells ($I_q$) which produce virions of the same strain.  Both infected cells and virions decay at a constant rate, in addition to infected cell death mediated by natural killer cells and effector CD8$^+$ T cells (also known as CTL), and virion neutralisation due to antibodies.  The model assumes that the total number of cells is constant, such that $\left(1-\frac{T+\sum_{q=1}^Q I_q}{T_0}\right)$ is the proportion of dead cells; target cells are then replenished at a rate proportional to the product of the number of uninfected cells and dead cells~\citep{Hancioglu2007}.  \citet{Cao2015} modelled three possible mechanisms for the innate immune response, but as the present study concentrates on the cellular adaptive immune response, we use one innate immune mechanism only (mechanism 3 in~\citet{Cao2015}), namely natural killer cells which kill infected cells.  We note that the qualitative results presented in this study are unchanged if a different innate immune mechanism is chosen (results not shown).  We assume that natural killer cells are present in proportion to type I interferon ($F$), which is produced in response to infected cells.  Effects of natural killer cells other than cytolysis (such as production of type II interferon) are not modelled, but may be included in future studies.

The model also captures the role of antibodies ($A_q$), responsible for strain-specific viral clearance and induction of long-term sterilising immunity (Fig.~\ref{fig:flow}B).  We assume that any cross-reactive humoral immunity between the strains plays a subdominant part in the immune response~\citep{Terajima2013,Grebe2008}, such that there is a one-to-one correspondence between antibodies and the strains upon which they act.  The equations which describe the production of antibodies are given by Eq.~\ref{eq:abs_multistrain}:

\begin{subequations}
\begin{align}
\frac{dB_{0q}}{dt} &= -\frac{V_q}{k_{Bq}+V_q} \beta_{Bq}B_{0q} \label{eq:B0}\\
\frac{dB_{1q}}{dt} &= \frac{V_q}{k_{Bq}+V_q} \beta_{Bq}B_{0q} - \frac{n_{Bq}B_{1q}}{\tau_{Bq}}- \delta_{Bq} B_{1q} \label{eq:B1}\\
\frac{dB_{iq}}{dt} &= \frac{n_{Bq}(2B_{i-1,q}-B_{iq})}{\tau_{Bq}} - \delta_{Bq} B_{iq}, \qquad i = 2, ..., n_B \label{eq:Bi}\\
\frac{dP_q}{dt} &= \frac{2n_{Bq}B_{n_B,q}}{\tau_{Bq}} - \delta_{Bq} P_{q} \\
\frac{dA_q}{dt} &= p_{Aq} P_{q} - \delta_{Aq}A_q. 
\end{align}
\label{eq:abs_multistrain}
\end{subequations}

The stimulation of naive B cells ($B_{0q}$) takes the form of a saturating function, as shown in Eq.~\ref{eq:B0}.  Once stimulated, naive B cells become plasmablasts ($B_{iq}$ where $i$ denotes the stage of plasmablast)~\citep{MingesWols2015}, as shown in Eq.~\ref{eq:B1}.  In Eqs.~\ref{eq:B1} and \ref{eq:Bi}, the plasmablasts undergo programmed proliferation for time $\tau_{Bq}$, passing through $n_{Bq}$ divisions/stages, until reaching the terminal stage -- plasma cells ($P_q$).  The proliferation process can alternatively be modelled using delay differential equations, where the equivalent number of plasma cells appears after some delay~\citep{Cao2016}; the two models give qualitatively similar results, but the ordinary differential equation approach here makes the divisions explicit.  Plasma cells ($P_q$) produce antibodies ($A_q$) which bind to virions and neutralise them.  Plasmablasts can also produce antibodies, but at a lower rate~\citep{MingesWols2015}, which we neglect in our model.

The cellular adaptive immune response, which is responsible for lysis of infected cells by effector CTL (Fig.~\ref{fig:flow}C), is described in Eqs.~\ref{eq:cd8_multistrain1} and  ~\ref{eq:cd8_multistrain2}.  Eq.~\ref{eq:cd8_multistrain1} describes the proliferation and differentiation of naive CD8$^+$ T cells into effector CTL and the subsequent formation of memory CD8$^+$ T cells.

\begin{subequations}
\begin{align}
\frac{dC_{j}}{dt} &= -\frac{\sum_{q=1}^Q I_q/k_{Cjq}}{1+\sum_{q=1}^Q I_q/k_{Cjq}} \beta_{Cj}C_{j}  \label{eq:dC0dt}\\
\frac{dE_{1j}}{dt} &= \frac{\sum_{q=1}^Q I_q/k_{Cjq}}{1+\sum_{q=1}^Q I_q/k_{Cjq}} \beta_{Cj}C_{j} - (\frac{n_{Ej}}{\tau_{Ej}}+ \delta_{Ej})E_{1j} \label{eq:E1}\\
\frac{dE_{ij}}{dt} &= \frac{n_{Ej}(2E_{i-1,j} - E_{ij})}{\tau_{Ej}} - \delta_{Ej} E_{ij}, \qquad i = 2,...,n_E-1 \label{eq:Ei}\\
\frac{dE_{n_{Ej}}}{dt} &= \frac{2n_{Ej}E_{n_E-1,j}}{\tau_{Ej}} - \delta_{Ej} E_{n_{Ej}} \label{eq:En}\\
E_j &= \sum_{i = 1}^{n_{E}}E_{ij}\\
\frac{dM_{j}}{dt} &= \epsilon_{j}\delta_{Ej} E_{n_{Ej}} +\hat{\epsilon}_{j}\delta_{\hat{E}j} \hat{E}_{n_{\hat{E}j}} - \delta_{Ej}M_{j} - \frac{M_{j}}{\tau_{Mj}}. 
  \end{align}
    \label{eq:cd8_multistrain1}
  \end{subequations}
  
We assume that there are $J$ pools of naive CD8$^+$ T cells, and that each of these recognises a single viral epitope, although it is possible for cells infected with different strains to present the same epitope.  The value for $J$ sufficient to model specific viruses is virus-dependent.  For example, if the cellular adaptive immune response is completely cross-reactive between strains, a single CD8$^+$ T cell pool ($J=1$) may be sufficient; for partial cross-reactivity, at least three CD8$^+$ T cell pools are required.

We assume that within each pool, all CD8$^+$ T cells are identical; the possibility that pools may consist of CD8$^+$ T cells with different responsiveness to the epitope is discussed in Section~\ref{sec:discussion}. Epitopes are presented to CD8$^+$ T cells by MHC class I molecules on the surface of either directly infected cells or on the surface of dendritic cells that have taken up the antigen and cross-presented it.  Cross-presentation can greatly influence the overall level of epitope presentation in influenza infection~\citep{Crowe2003}, but under a common modelling assumption (see e.g.~\citet{Chao2004}) that both direct presentation and cross-presentation are proportional to the number of infected cells, the model need not explicitly model cross-presentation. 

Naive CD8$^+$ T cell pool $j$ ($C_j$) is stimulated by interaction with the pMHC, as shown in Eq.~\ref{eq:dC0dt}; the stimulation function is a saturating function, such that $k_{Cjq}$ is the number of cells infected with strain $q$ required for (direct- and cross-) presentation of epitopes which yields half-maximal stimulation of the cellular adaptive immune response~\citep{DeBoer2001,Davenport2002,Chao2004}.   Once stimulated, naive CD8$^+$ T cells divide to become effector CTL of the first stage ($E_{1j}$), as shown in Eq.~\ref{eq:E1}; the effector cells then undergo programmed proliferation~\citep{Kaech2001,vanStipdonk2001,Wodarz2005} for time $\tau_{Ej}$, passing through $n_{Ej}$ divisions, as shown in Eqs.~\ref{eq:E1}, \ref{eq:Ei} and \ref{eq:En}.  The subscript $i$ denotes the stage of effector CTL ($1$ to $n_{Ej}$).

When effector CTL reach their last stage (the dynamics of which are described by Eq.~\ref{eq:En}), they no longer divide; instead, some fraction $\epsilon_j$ survive to become memory CD8$^+$ T cells ($M_j$).  These memory CD8$^+$ T cells are refractory in the sense that they cannot be restimulated for some time~\citep{Kaech2002}, which we model as exponentially distributed with mean $\tau_{Mj}$.  After this time, the memory CD8$^+$ T cells can be restimulated, and we relabel them $\hat{C}_j$, where $(\hat{\quad})$ indicates memory.  Upon restimulation, memory CD8$^+$ T cells once again become effector CTL $\hat{E}_j$, which can become refractory memory CD8$^+$ T cells $M_j$; the process is the same as in Eq.~\ref{eq:cd8_multistrain1}.

  \begin{subequations}
  \begin{align}
\frac{d\hat{C}_{j}}{dt} &= \frac{M_{j}}{\tau_{Mj}} - \frac{\sum_{q=1}^Q I_q/k_{\hat{C}jq}}{1+\sum_{q=1}^Q I_q/k_{\hat{C}jq}} \beta_{\hat{C}jq}\hat{C}_{j}\label{eq:dC0hatdt}\\
\frac{d\hat{E}_{1j}}{dt} &= \frac{\sum_{q=1}^Q I_q/k_{\hat{C}jq}}{1+\sum_{q=1}^Q I_q/k_{\hat{C}jq}} \beta_{\hat{C}jq}\hat{C}_{j} - (\frac{n_{\hat{E}j}}{\tau_{\hat{E}j}}+ \delta_{\hat{E}j})E_{1\hat{j}}\\
\frac{d\hat{E}_{ij}}{dt} &= \frac{n_{\hat{E}j}(2\hat{E}_{i-1,j} - \hat{E}_{ij})}{\tau_{\hat{E}j}} - \delta_{\hat{E}j} \hat{E}_{ij}, i = 2,...,n_{\hat{E}}-1\\
\frac{d\hat{E}_{n_{\hat{E}j}}}{dt} &= \frac{2n_{\hat{E}j}\hat{E}_{n_{\hat{E}}-1,j}}{\tau_{\hat{E}j}} - \delta_{\hat{E}j} \hat{E}_{n_{\hat{E}j}} \\
\hat{E}_j &= \sum_{i = 1}^{n_{\hat{E}}}\hat{E}_{ij}. 
  \end{align}
  \label{eq:cd8_multistrain2}
  \end{subequations}

All effector CTL, $E$ and $\hat{E}$, kill infected cells according to the law of mass action~\citep{Nowak1996}, at rates $\kappa_{Ejq}$ and $\kappa_{\hat{E}jq}$ respectively. Unlike antibodies, effector CTL are not lost during the binding and killing process.  

\begin{figure}[t]
\centering
\includegraphics[width = \textwidth]{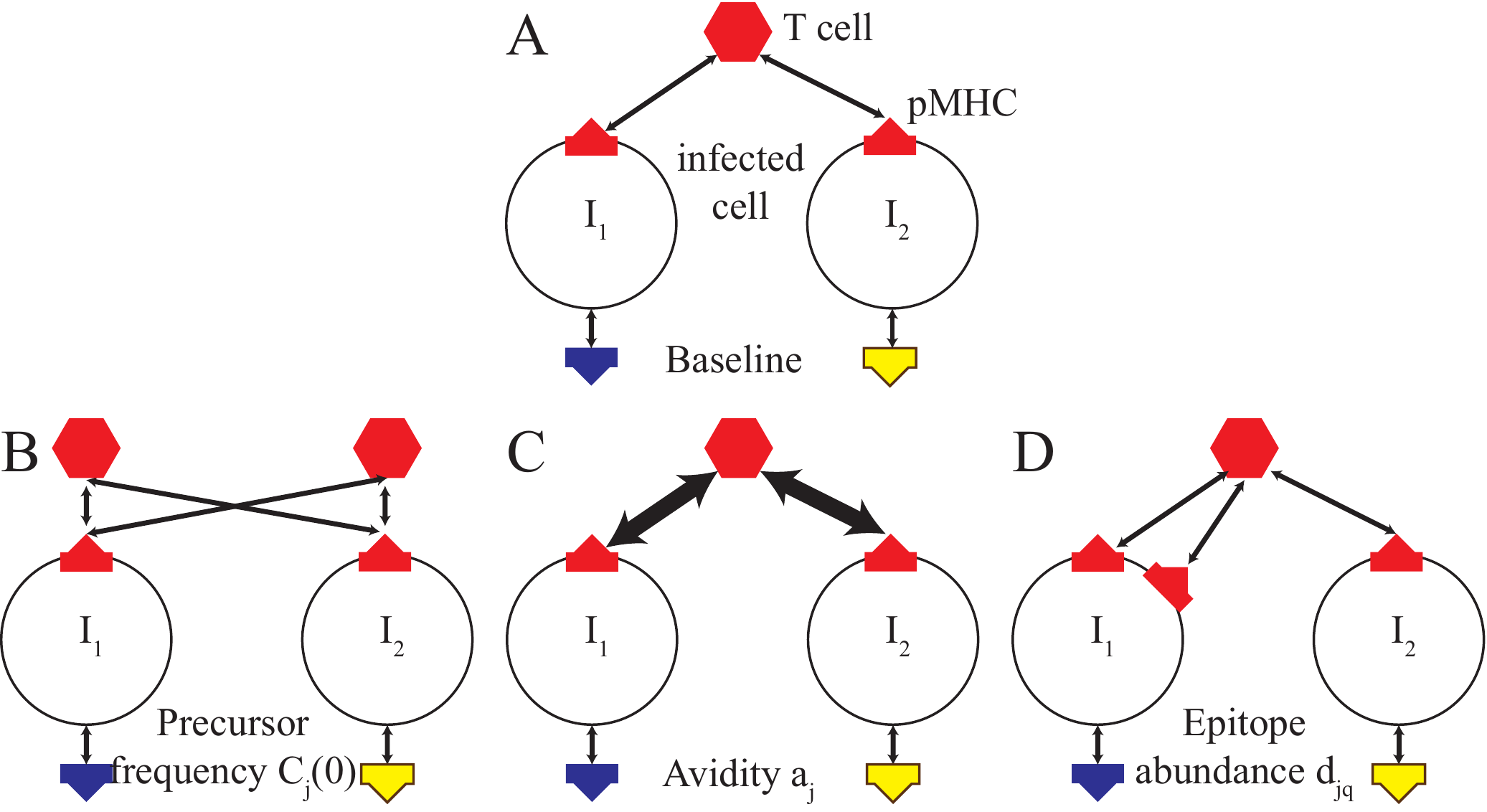}
\caption{Changing the strength of the cross-reactive cellular adaptive immune response.  Circles represent cells infected with strains 1 and 2; triangles on top of rectangles represent pMHC complexes; hexagons represent CD8$^+$ T cells.  The red epitope is shared between strains while the blue and yellow epitopes are not.  Relative to the baseline case (A), we can change the number of infected cells required for half-maximal stimulation of naive/memory CD8$^+$ T cells ($k_{Cjq}$,$k_{\hat{C}jq}$) and the lysing rate of infected cells by effector CTL ($\kappa_{Ejq}$, $\kappa_{\hat{E}jq}$).  This is accomplished by changing (B) the initial number (precursor frequency) of naive CD8$^+$ T cells $C_{j}(0)$; (C) the avidity of the interaction between the CD8$^+$ T cell and the pMHC complex which presents the epitope $a_j$; or (D) the epitope abundance (the number of pMHC complexes on the surface of the cell) $d_{jq}$.}
\label{fig:cartoon}
\end{figure}

\begin{figure}[htbp]
\centering
\includegraphics[width =\textwidth]{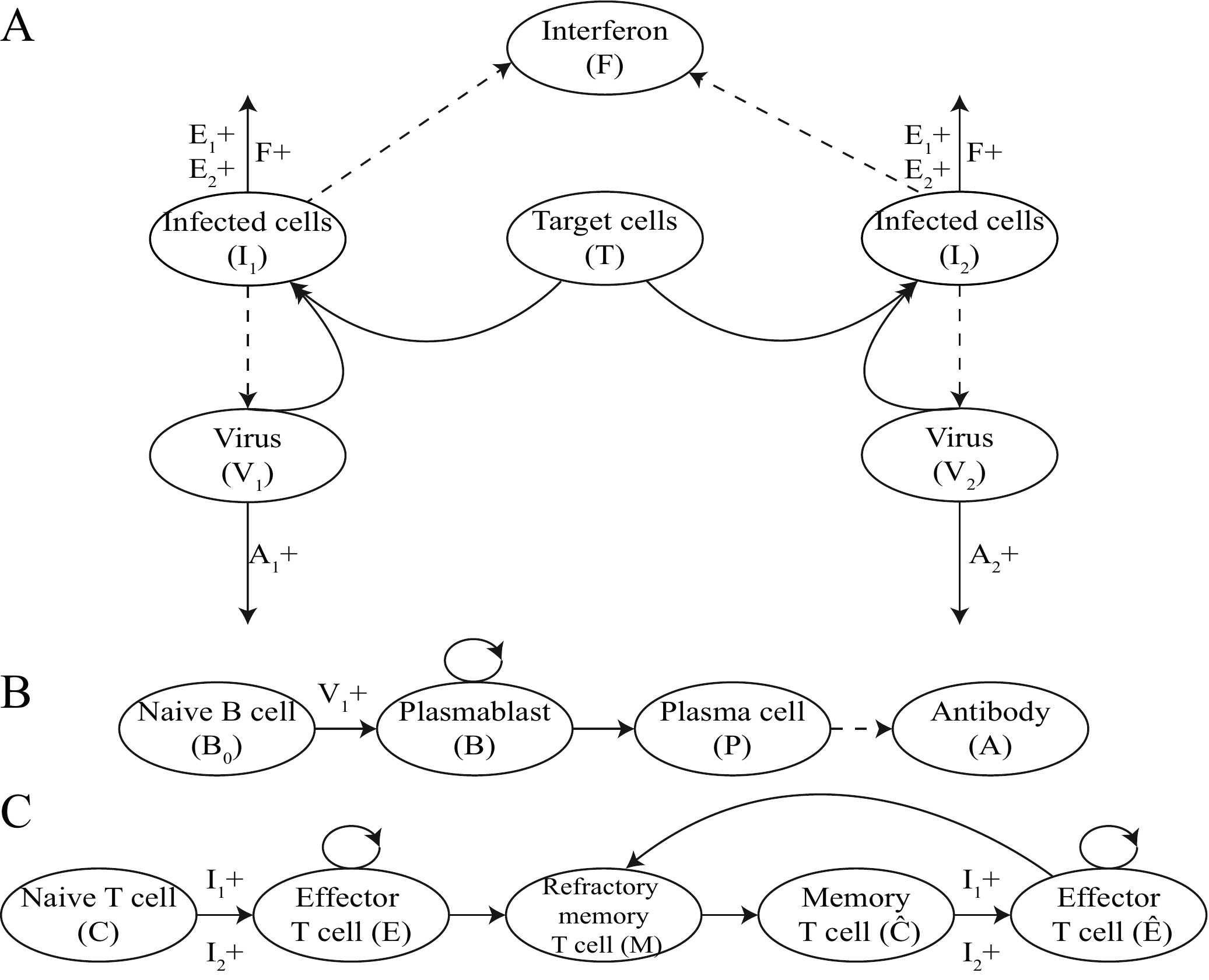}
\caption{The within-host influenza model illustrated using two strains and one CD8$^+$ T cell pool.  (A) Viral dynamics and innate immune response; (B) Humoral adaptive immune response (shown for virus 1 only; an identical, independent set of compartments exist for virus 2); (C) Cellular adaptive immune response.  Solid arrows indicate transitions between compartments or death (shown only for immune-enhanced death processes); dashed arrows indicate production; plus signs indicate an increased transition rate due to the indicated compartment.}
\label{fig:flow}
\end{figure}

The number of cells infected with strain $q$ required for half-maximal stimulation of naive CD8$^+$ T cells from pool $j$, $k_{Cjq}$, and the lysing rate of cells infected with strain $q$ by effector CTL from pool $j$, $\kappa_{Ejq}$, are both measures of cross-reactivity.  We model $k_{Cjq}$ and $\kappa_{Ejq}$ as functions of the avidity of the T cell-pMHC interaction, $a_j$, and the epitope abundance per infected cell, $d_{jq}$ (see Fig.~\ref{fig:cartoon}); these will be elaborated upon in Section~\ref{sec:cross}.  We assume that the killing rate is directly correlated with both avidity and epitope abundance, while the number of infected cells required for half-maximal stimulation is inversely correlated with avidity and epitope abundance.  Hence,

\begin{subequations}
\begin{align}
k_{Cjq} &= \frac{\tilde{k}_C}{a_jd_{jq}}\\
\kappa_{Ejq} &= \tilde{\kappa}_Ea_jd_{jq}
\label{eq:def_cross}
\end{align}
\end{subequations}
where $\tilde{k}_C$ and $\tilde{\kappa}_E$ are baseline values (similar equations hold for the memory ${(\enskip\hat{}\enskip)}$ versions of these parameters).

\subsection{Model parameters and solution}

In Section \ref{sec:cross}, to model immune responses of different strengths, we vary $k_C$ and $k_{\hat{C}}$, the number of infected cells for half-maximal stimulation of naive and memory CD8$^+$ T cells respectively.  We will also vary $\kappa_E$ and $\kappa_{\hat{E}}$, the lysing rate of infected cells by effector CTL which originate from naive and memory CD8$^+$ T cells respectively.  All other parameters are chosen to be identical for all strains. These parameters are listed in Tables~\ref{table:parameters},~\ref{table:parameters_abs} and~\ref{table:parameters_cd8}. The initial values which are the same between the strains are listed in Table~\ref{table:iv}.  In addition, we define a baseline value $\tilde{C}(0)$ for the initial number of naive CD8$^+$ T cells which respond to a particular epitope. 

Given the outstanding immunological uncertainties surrounding the differences between naive and memory CD8$^+$ T cells (see Section~\ref{sec:discussion}), we assume that memory CD8$^+$ T cells $\hat{C}$ are identical to naive CD8$^+$ T cells $C$, such that the effects of memory are purely due to an increased number of CD8$^+$ T cells after a primary infection. In other words, we set the parameters $k_C$, $\kappa_E$, $\beta_C$, $n_{E}$, $\tau_E$ and $\delta_E$ to be equal to $k_{\hat{C}}$, $\kappa_{\hat{E}}$, $\beta_{\hat{C}}$, $n_{\hat{E}}$, $\tau_{\hat{E}}$ and $\delta_{\hat{E}}$.  In consequence, the change between a primary and secondary infection is due to initial conditions: $\hat{C}$ after a primary infection is greater than $C(0)$, which is true as long as the number of divisions for effector CTL and the proportion of effector CTL which become memory CD8$^+$ T cells is not too low, and the death rate of effector CTL is not too high (results not shown).

Because we are interested in the qualitative behaviour of the model, most parameters are chosen based on existing literature.  Of these, a large number are taken from \citet{Cao2016}, where a similar model for single-strain infection was calibrated to viral load, antibody and CTL data; we use these model-derived parameters for our study.  Other parameters are chosen such that the time course of a single infection fits the following criteria based on experimental observations:

\begin{itemize}
\item The peak viral load occurs at about 2 days post-infection~\citep{Laurie2015}
\item The innate immune response is most active 2 \textendash 7 days post-infection~\citep{Carolan2015,Pawelek2012}
\item Antibodies appear after 5 days post-infection and peak at about 20 days post-infection~\citep{Miao2010}
\item CD8$^+$ T cells peak at about 8 days post-infection~\citep{Kaech2002}
\item When both humoral and cellular adaptive immune responses are removed, chronic infection occurs~\citep{Iwasaki1977,Kris1988}
\item When the cellular adaptive immune response is removed, resolution of the infection is delayed~\citep{Hou1992}
\end{itemize}

The ordinary differential equations are solved using MatlabR2014b's \textit{ode15s}, with default integration settings.  To avoid infections rebounding from unrealistically low numbers of virions/infected cells, infections are truncated by setting $V = I = 0$ when both the number of virions and infected cells drops below 1.

Code to reproduce all of the figures in the study is available at \url{https://ada_yan@bitbucket.org/ada_yan/cross-reactivity.git}.

\section{Results}
\label{sec:results}

\begin{figure}[htbp]
\centering
\includegraphics[width = \textwidth]{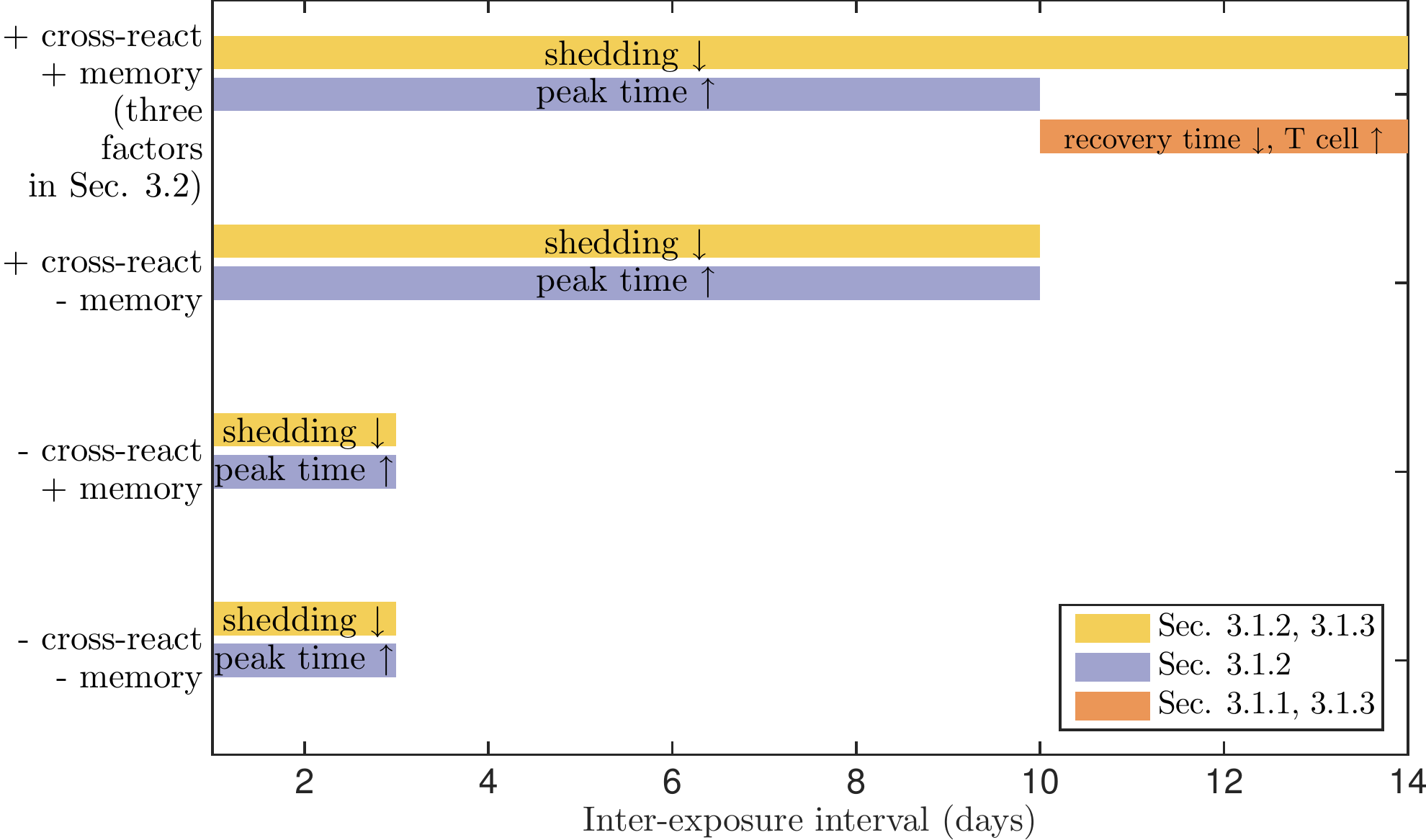}
\caption{Summary of the effect of the inter-exposure interval, cross-reactivity and memory on the outcome of the second infection.  The bars indicate the inter-exposure intervals for which the labelled effects are observed, for each scenario on the vertical axis.  The scenarios are (from top) (1) cross-reactivity between the two strains and memory CD8$^+$ T cells present; (2) cross-reactivity but no memory CD8$^+$ T cells; (3) no cross-reactivity but memory CD8$^+$ T cells present; (4) no cross-reactivity or memory CD8$^+$ T cells.  The yellow bar corresponds to a decrease in shedding; the violet bar corresponds to a delay in the time to peak viral load relative to a primary infection;  the orange bar corresponds to a decrease in recovery time, as well as an increase in the number of CD8+ T cells after the second infection relative to the first.  The same colours are used for the axis labels of the relevant quantities in Figs.~\ref{fig:T} and \ref{fig:stats}.  For short inter-exposure intervals (1--3 days), reduced shedding and delayed infection are due to the innate immune response, and are thus independent of whether cross-reactivity and/or memory are present.  For medium inter-exposure intervals (5--10 days), cross-reactivity is required to reduce shedding and delay infection.  For long inter-exposure intervals (10--14 days), only the model with cross-reactivity and memory reduces the recovery time and boosts the number of CD8+ T cells.}
\label{fig:summary}
\end{figure}

Before presenting detailed results, we first outline the main results of the paper.  The effect of a cross-reactive immune response which allows for effector CTL to differentiate into memory CD8$^+$ T cells depends on the interval between exposures to the two viruses.  For short inter-exposure intervals (1--3 days), the innate immune response delays a second infection and reduces shedding relative to a first infection, independent of whether cross-reactivity and/or memory are present in the cellular adaptive immune response (Fig. 3).  For medium inter-exposure intervals (5--10 days), delayed infection and reduced shedding are instead due to the cellular adaptive immune response, such that cross-reactivity is required to delay infection and reduce shedding.  For long inter-exposure intervals (10--14 days), both cross-reactivity and memory are required to decrease the recovery time of the second infection relative to the first, and to boost the number of CD8$^+$ T cells after a second infection to numbers much larger than for a first infection.  The inter-exposure intervals for which delaying of a second infection, reduced shedding, reduced recovery time and boosting of CD8$^+$ T cell numbers are observed for the four scenarios explored in this section are summarised in Fig.~\ref{fig:summary}.  We will now discuss each of these effects in detail.

\subsection{The effect of the inter-exposure interval on the outcome of a second infection depends on the cross-reactivity between strains}\label{sec:iei}

In this section, we compare scenarios where there is cross-reactivity between two strains to scenarios where there is no cross-reactivity.  The former is modelled using a single CD8$^+$ T cell pool which equally responds to the two strains.  The parameters are $k_{C11} = k_{C12} = k_{\hat{C}11} = k_{\hat{C}12} = \tilde{k}_{C}$ and $\kappa_{E11} = \kappa_{E12} = \kappa_{\hat{E}11} = \kappa_{\hat{E}12} = \tilde{\kappa}_{E}$; the initial number of naive CD8$^+$ T cells is $C_1(0) =  \tilde{C}(0)$.  The latter is modelled using two CD8$^+$ T cell pools, each of which only responds to one strain.  The parameters are $k_{C11} = k_{C22} = k_{\hat{C}11} = k_{\hat{C}22} = \tilde{k}_{C}$ and $\kappa_{E11} = \kappa_{E22} = \tilde{\kappa}_E$; $k_{C12} = k_{C21} = k_{\hat{C}12} = k_{\hat{C}21} = \infty$ and $\kappa_{E12} = \kappa_{E21} = \kappa_{\hat{E}12} = \kappa_{\hat{E}21} = 0$; the initial number of naive CD8$^+$ T cells is $C_1(0) = C_2(0) = \tilde{C}(0)$.  For each scenario (presence or absence of cross-reactivity), we compare the sub-cases where effector CTL can become memory CD8$^+$ T cells and where effector CTL decay without becoming memory CD8$^+$ T cells.  The latter is modelled by setting $\epsilon_E = 0$.

\begin{figure}[p]
\centering
\includegraphics[width =\textwidth]{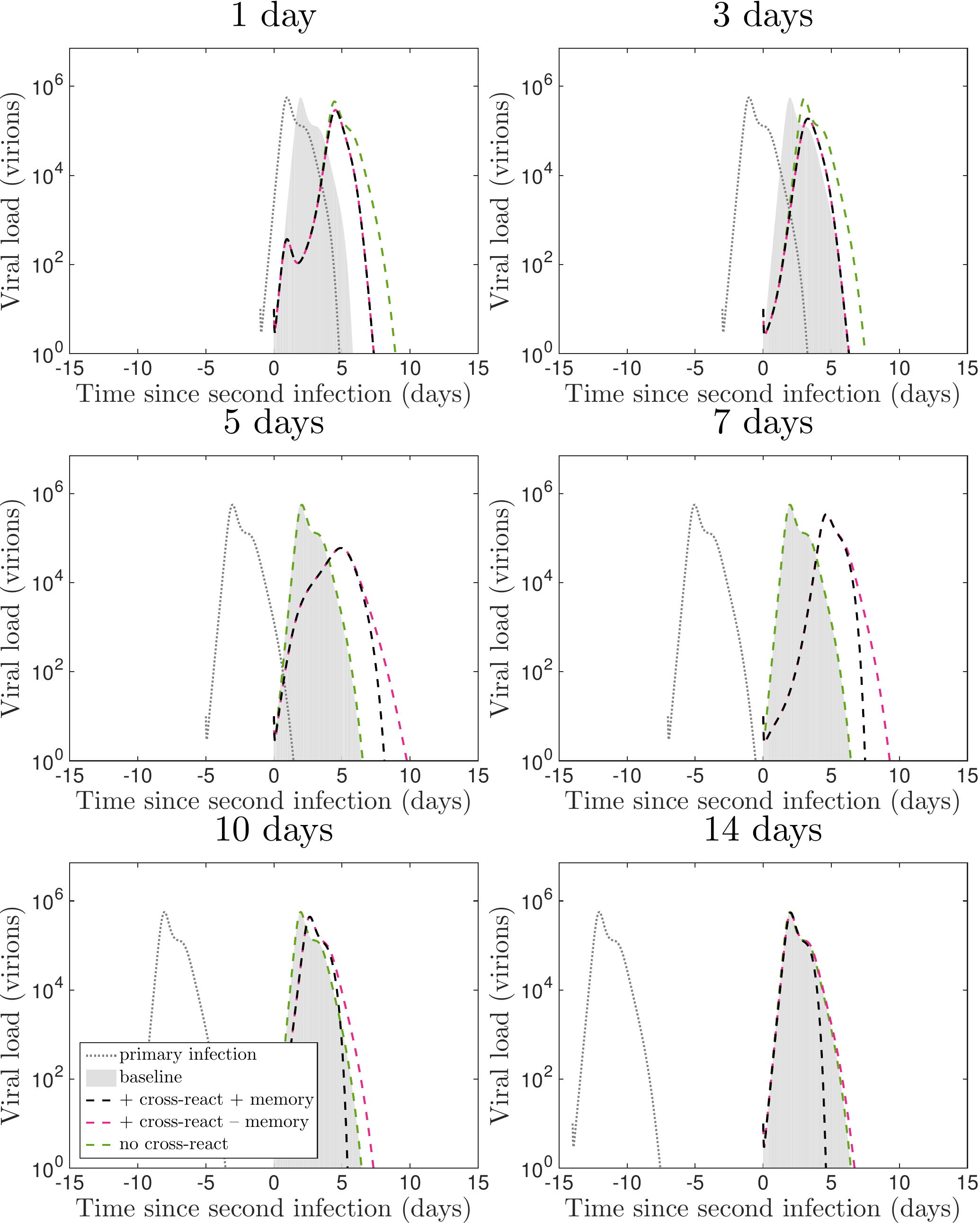}
\caption{Viral load for a second infection for three cases: (1) cross-reactivity between the two strains and memory CD8$^+$ T cells present (black dashed line); (2) cross-reactivity but no memory CD8$^+$ T cells (magenta dashed line); (3) no cross-reactivity between the strains (green dashed line).  Note: in the case where there is no cross-reactivity between strains, the time course of the second infection is not changed by the presence of memory CD8$^+$ T cells.  These are compared to what the viral load would have been had the primary infection been absent (grey area).  The grey dotted line indicates the viral load for the primary infection.  Captions indicate inter-exposure intervals.}
\label{fig:viral_load}
\end{figure}

\begin{figure}[htbp]
\centering
\includegraphics[width =\textwidth]{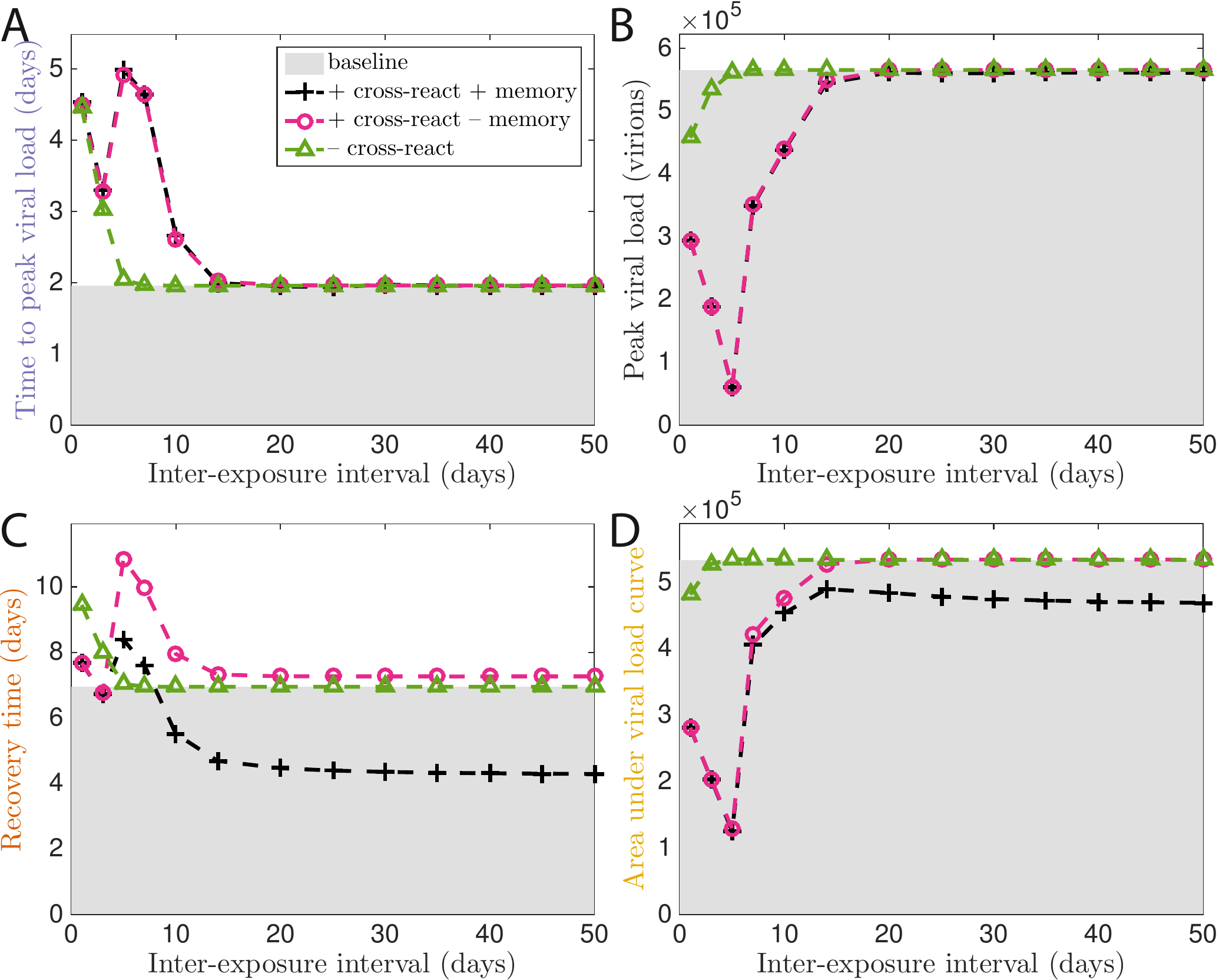}
\caption{The effect of changing the inter-exposure interval on (A) the time to peak viral load (B) the peak viral load (C) the recovery time (D) the area under the viral load curve for the second infection.  We show the results for three cases: (1) cross-reactivity between the two strains and memory CD8$^+$ T cells present (black line with crosses); (2) cross-reactivity but no memory CD8$^+$ T cells (magenta line with circles); (3) no cross-reactivity between the strains (green line with triangles).  These are compared to the baseline values obtained from a single infection (shaded area).  The recovery time is defined as the time from inoculation to when both the number of infected cells and the number of virions drops below 1.  Note: our definition of recovery time is different to that by \citet{Laurie2015}, where the recovery time was defined as the duration for which the second virus exceeds an experimental threshold.}
\label{fig:stats}
\end{figure}

Figure~\ref{fig:viral_load} shows the viral load for two successive infections under three different sets of assumptions, as detailed in the figure captions.  For short inter-exposure intervals (1--3 days), the innate immune response delays viral kinetics under all three sets of assumptions.  For medium inter-exposure intervals (5--10 days), cross-reactivity further delays a second infection and reduces shedding.  For long inter-exposure intervals (10--14 days), both cross-reactivity and memory are needed to shorten the second infection.  Figure~\ref{fig:stats} shows the summary statistics for Fig.~\ref{fig:viral_load}, which we will now analyse.

\subsubsection{For medium inter-exposure intervals, cross-reactivity further delays a second infection and reduces shedding}\label{sec:medium_iei}

For medium inter-exposure intervals (5--10 days), the time to peak viral load is delayed when there is cross-reactivity between the strains, due to the presence of cross-reactive effector CTL during the exponential growth phase of the second infection.  The growth rate is also slowed down, such that the viral load reaches a smaller value before the adaptive immune response clears the infection (Fig.~\ref{fig:stats}B), resulting in a smaller area under the viral load curve (Fig.~\ref{fig:stats}D).  The violet and yellow bars in Fig.~\ref{fig:summary} indicate that when the cellular adaptive immune response is cross-reactive, the second infection is delayed and shedding is reduced for a larger range of inter-exposure intervals.

\subsubsection{For long inter-exposure intervals, cross-reactivity and memory shorten the recovery time for a second infection and reduce shedding}\label{sec:long_iei}

When there is no cross-reactivity, for long inter-exposure intervals ($> 10$ days), the recovery time for the second infection approaches that of a primary infection (Fig.~\ref{fig:stats}C).  This is due to the temporary nature of the innate immune response, and the cellular adaptive immune response approaching a steady state. However, when cross-reactivity and memory CD8$^+$ T cells are present, the recovery time is shortened relative to that for a primary infection (Fig.~\ref{fig:stats}C).  This is consistent with the experimental observations of \citet{Laurie2015} and \citet{Christensen2000}.    The area under the viral load curve is also reduced (Fig.~\ref{fig:stats}D).  These results are indicated by the orange bar in Fig.~\ref{fig:summary}.  Under our model where memory cells are functionally identical to naive CD8$^+$ T cells, the advantage is conferred by having extra memory CD8$^+$ T cells, such that the number of effector CTL is higher for a second infection, and the lysis rate of infected cells $\sum_{j=1}^J (\kappa_{Ejq}E_j + \kappa_{\hat{E}jq}\hat{E}_j)$ in Eq. \ref{eq:dIdt} increases.  

In the absence of memory CD8$^+$ T cells (but when cross-reactivity is present), the recovery time for the second infection is in fact slightly longer than for the first infection, due to the partial depletion of naive CD8$^+$ T cells by the first infection.  This confirms the crucial role of memory CD8$^+$ T cells in shortening the second infection.  At initiation of the second infection, prior to recall of memory CD8$^+$ T cells, the virus is able to replicate effectively.  Then upon recall of memory CD8$^+$ T cells, effector CTL levels grow to higher levels than the first infection, resulting in faster clearance.  For this to be due to residual effector CTL from the first infection, the time at which the number of effector CTL peaks would have to be longer than the inter-exposure interval; as CTL peak at about eight days post-exposure~\citep{Kaech2002}, this is not true for longer inter-exposure intervals.  Hence, it is more plausible that a second growth phase of effector CTL occurs following recall of memory CD8$^+$ T cells.

The result that for long inter-exposure intervals, cross-reactivity and memory are required to shorten the recovery time for a second infection is robust to changes in model parameters.  A sensitivity analysis for Fig.~\ref{fig:stats}C is presented in Fig.~\ref{fig:sensitivity_recovery_time_iei} in \ref{sec:duration_sensitivity}.  The parameters varied are $p_V$, the production rate of virions from cells infected with either virus; $p_F$, the production rate of interferon from cells infected with either virus; and $\kappa_A$, the neutralising rate of virions of either strain by antibodies.  These parameters vary viral replication (in the absence of the immune response), the innate immune response, and the humoral adaptive immune response respectively.  Across biologically plausible ranges of these parameters, both cross-reactivity and memory are required to shorten the recovery time for a second infection.

\subsubsection{A second infection boosts \texorpdfstring{CD8$^+$}{CD8+} T cell numbers when cross-reactivity and memory are present}\label{sec:boost}

\begin{figure}[t]
\centering
\includegraphics[width = \textwidth]{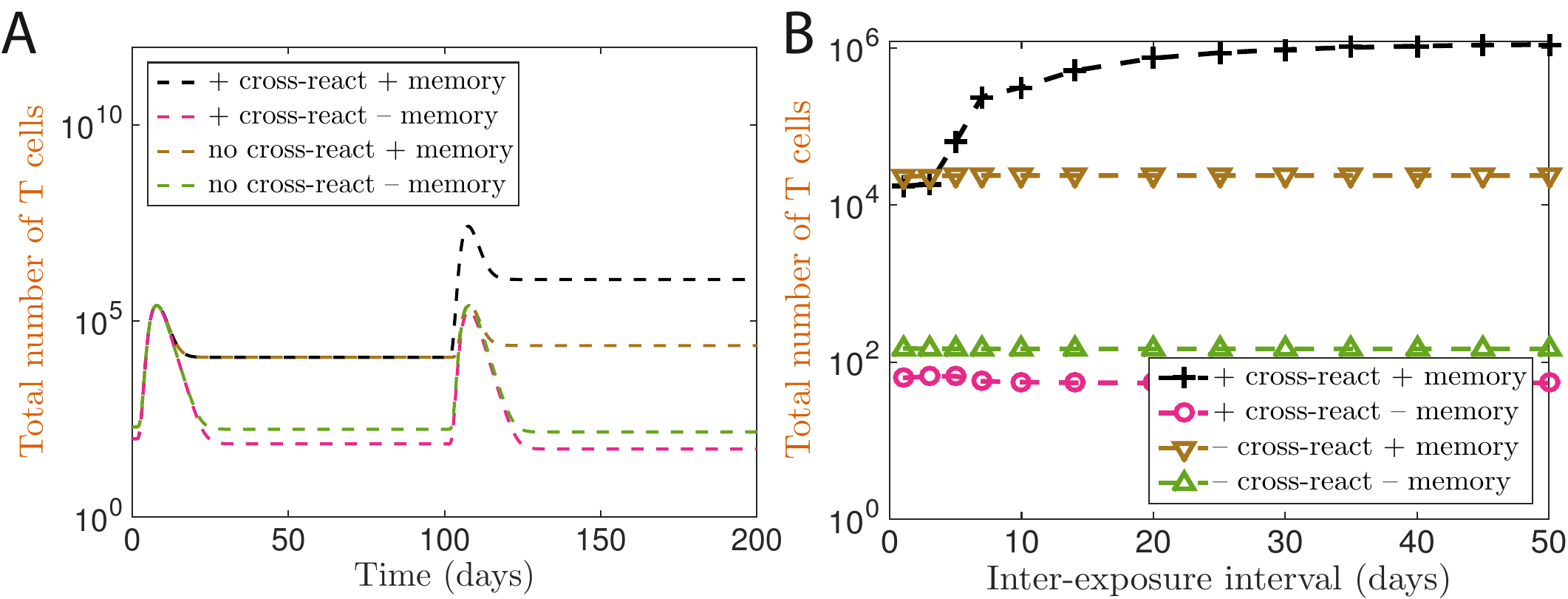}
\caption{(A) Time course of the total number of CD8$^+$ T cells for a 100-day inter-exposure interval for four cases: (1) cross-reactivity between the two strains and memory CD8$^+$ T cells present (black line); (2) cross-reactivity but no memory CD8$^+$ T cells (magenta line); (3) no cross-reactivity but memory CD8$^+$ T cells (brown line); (4) no cross-reactivity or memory CD8$^+$ T cells (green line).  Note: for the cases without memory CD8$^+$ T cells, the difference between the cross-reactive and non-cross-reactive case is due to the choice of initial conditions, whereby the number of CD8$^+$ T cells in each pool is kept constant, rather than the total number of CD8$^+$ T cells.  (B) The total number of CD8$^+$ T cells 100 days after the second infection, varying the inter-exposure interval.  The total number of CD8$^+$ T cells is defined as $\sum_{j=1}^J C_j + E_j + M_j + \hat{C}_j + \hat{E}_j$.}
\label{fig:T}
\end{figure}

When the model does not include memory CD8$^+$ T cells, the total number of CD8$^+$ T cells is slightly less after an infection than before, due to the decay of the CD8$^+$ T cells which responded to the infection (Fig.~\ref{fig:T}A).  On the other hand, when the model includes effector CTL which can become memory CD8$^+$ T cells, infection greatly boosts the number of CD8$^+$ T cells. When there is no cross-reactivity between the strains, a separate pool of CD8$^+$ T cells expands and contracts upon the second infection, such that the final number of cells is roughly twice that for a single infection.  On the other hand, when there is cross-reactivity between the strains, memory cells generated by the first infection are restimulated, so the CD8$^+$ T cell response to the second infection is much larger (two orders of magnitude larger than for a single infection).  

In this case, as the inter-exposure interval increases, the number of CD8$^+$ T cells after the second infection increases (Fig.~\ref{fig:T}B) -- the larger the inter-exposure interval, the more effector CTL from the first infection have differentiated into memory CD8$^+$ T cells, and so the larger the pool of memory CD8$^+$ T cells which can be induced to proliferate by a second infection.  As indicated by the orange bar in Fig.~\ref{fig:summary}, the number of CD8$^+$ T cells after the second infection is at least an order of magnitude larger than for the first infection if (and only if) both cross-reactivity and memory are present, for inter-exposure intervals 10 days or longer.  

For long inter-exposure intervals, the precise duration of the inter-exposure interval becomes less important because the system approaches a steady state after resolution of the first infection.  The total number of T cells approaches this maximum if the inter-exposure interval is long enough such that by the second infection, most of the CD8$^+$ T cells stimulated by the first infection have differentiated into memory cells which are capable of being restimulated.  This time roughly corresponds to the sum of the mean time until naive CD8$^+$ T cell stimulation, the mean total time for effector CD8$^+$ T cell expansion, the mean time that effector CD8$^+$ T cells spend in their final stage, and the mean refractory period of memory CD8$^+$ T cells.  

The result that the final number of CD8$^+$ T cells after two heterologous infections is greater when there is cross-reactivity between the two strains is robust to changes in model parameters, as shown in Fig.~\ref{fig:sensitivity_total_T} in ~\ref{sec:total_T_sensitivity}.

\subsection{Increasing the strength of the cross-reactive response decreases the recovery time for the second infection, and increases the final number of \texorpdfstring{CD8$^+$}{CD8+} T cells}\label{sec:cross}

In the previous section, in the presence of memory CD8$^+$ T cells, we looked at two simplifying cases: 

\begin{enumerate}
\item Where a single shared epitope is immunodominant for two strains; and
\item Where the immunodominant epitopes for two strains are distinct (i.e. not shared), resulting in no cross-reactivity.
\end{enumerate}    

However, the overall strength of the cellular adaptive immune response to a single strain is the sum of the contributions from all of its epitopes.  In the context of protection against subsequent heterologous infection, we are therefore interested in the immune response to shared epitopes, as the effects of non-shared epitopes are limited to a single infection.

Accordingly, we now consider the effect of three immunological factors which change the strength of the immune response for the shared epitope.  We model variation in  

\begin{enumerate}
\item The initial naive T cell number, also known as the precursor frequency ($C_{j}(0)$);
\item The avidity of the pMHC-T cell interaction ($a_j$); and
\item The epitope abundance per cell infected with either the first or second virus ($d_{jq}$).
\end{enumerate}

These immunological concepts are illustrated in Fig.~\ref{fig:cartoon}.

These parameters can also be changed for the non-shared epitopes, but this only affects infection with a single strain.  To focus on the effect of changing the immune response to the shared epitope, we examine the case where there is a single immunodominant epitope which is shared between the strains (i.e. $J= 1$), and omit the non-shared epitopes.

\subsubsection{Factor 1: The precursor \texorpdfstring{CD8$^+$}{CD8+} T cell frequency}\label{sec:precursor_frequency}

\begin{figure}[t]
\centering
\includegraphics[width = \textwidth]{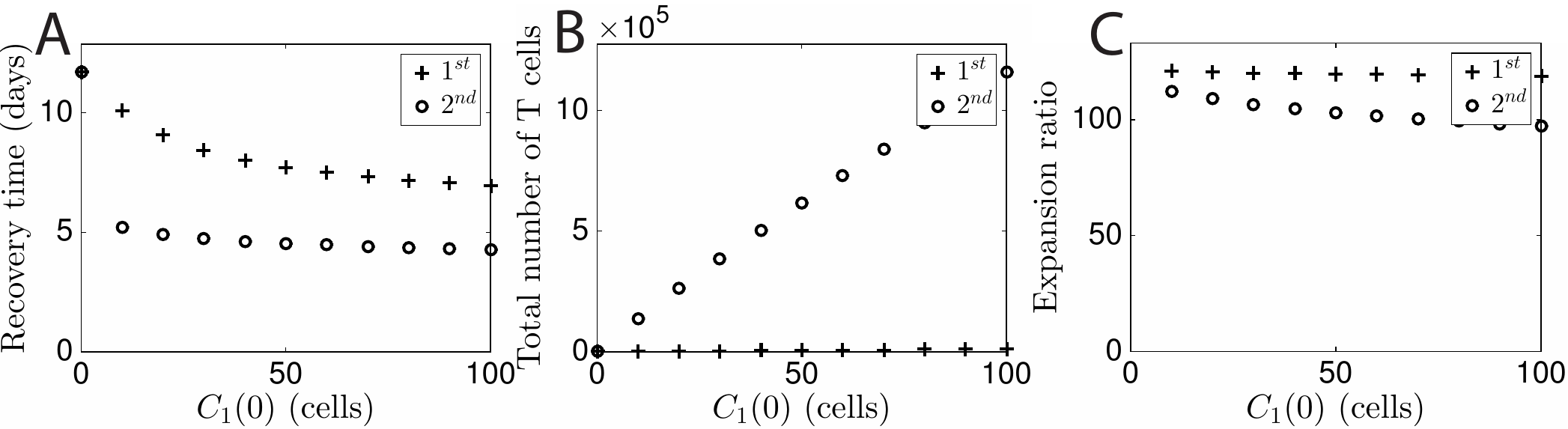}
\caption{The effect of changing the precursor frequency $C_1(0)$ on (A) the recovery time, (B) the total number of CD8$^+$ T cells 100 days after infection and (C) the expansion ratio, for the first and second infection.  The total number of CD8$^+$ T cells is defined as $C_1 + E_1 + M_1 + \hat{C}_1 + \hat{E}_1$.  The expansion ratio is the ratio between the number of CD8$^+$ T cells 100 days after an infection, and the number of CD8$^+$ T cells just before the infection.  The inter-exposure interval is fixed at 100 days, chosen to ensure that the number of CD8$^+$ T cells has settled to a steady state after the first infection. Results are robust for inter-exposure intervals greater than 20 days (not shown).}
\label{fig:change_precursor_frequency}
\end{figure}

We model the situation where the epitope abundance per infected cell is the same for both strains, and we vary the initial number of naive CD8$^+$ T cells $C_1(0)$ which respond to the epitope.  The avidity of the pMHC-T cell interaction is kept constant, i.e. $a_1 = 1$, $d_{11} = d_{12} = 1$ (Fig.~\ref{fig:cartoon}B).

As the precursor frequency $C_1(0)$ increases, the recovery time for the first infection decreases in an exponential-like manner (Fig.~\ref{fig:change_precursor_frequency}A), in agreement with the findings of \citet{Cao2016}.  The recovery time for the second infection also decreases in an exponential-like manner.  The decrease in recovery time is initially greater for the second infection, but levels out as the number of precursor CD8$^+$ T cells increases even further, because the time required for proliferation and differentiation of effector cells becomes a limiting factor.  The recovery time for the second infection is always shorter than that for the first infection. 

For such a long inter-exposure interval, the innate immune response has subsided by the time of the second infection, so the only difference between the initial conditions for the two infections is the number of CD8$^+$ T cells available for proliferation and differentiation into effector CTL.

As the number of precursor CD8$^+$ T cells increases, the total number of CD8$^+$ T cells after the second infection increases sublinearly (Fig.~\ref{fig:change_precursor_frequency}B).   This can be analysed in terms of the expansion ratio (Fig.~\ref{fig:change_precursor_frequency}C), defined as the number of CD8$^+$ T cells 100 days after an infection divided by the number of CD8$^+$ T cells before the infection. For example, when there are initially 10 CD8$^+$ T cells, each CD8$^+$ T cell results in approximately 120 CD8$^+$ T cells at the end of the first infection; after the second infection, each of these progeny result in approximately 110 CD8$^+$ T cells (i.e. the total expansion ratio is $120 \times 110 \approx 10^4$).  The expansion ratio is less for the second infection because the shortened infection decreases the duration and degree of CD8$^+$ T cell stimulation by infected cells, such that a smaller proportion of initial CD8$^+$ T cells proliferate and differentiate.  This result suggests that the boosting of the immune response by successive infections yields diminishing returns in terms of the proportion of new CD8$^+$ T cells, if not their absolute number.  This is consistent with the experimental results of \citet{Christensen2000}.

\subsubsection{Factor 2: The avidity of the pMHC-T cell interaction}\label{sec:avidity}

\begin{figure}[t]
\centering
\includegraphics[width = \textwidth]{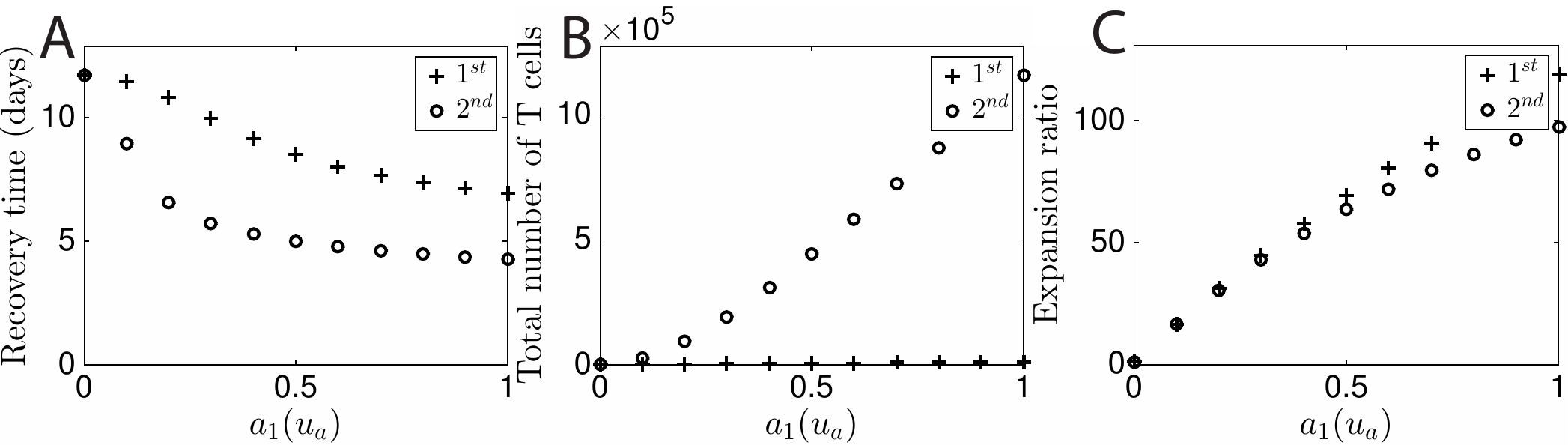}
\caption{The effect of changing the avidity $a_1$ on (A) the recovery time, (B) the total number of CD8$^+$ T cells 100 days after infection and (C) the expansion ratio, for the first and second infection.}
\label{fig:change_avidity}
\end{figure}

We now model the situation where the epitope abundance per infected cell is the same for both strains, but we vary the avidity of the pMHC-T cell interaction $a_1$.  The initial number of naive CD8$^+$ T cells which respond to the epitope is kept constant, i.e. $d_{11}  = d_{12} = 1$ and $C_1(0) = \tilde{C}(0)$ (Fig.\ref{fig:cartoon}C).

As the avidity $a_1$ increases, the recovery time for both the first and second infections decrease (Fig.~\ref{fig:change_avidity}A), in a similar manner to Fig.~\ref{fig:change_precursor_frequency}A.

The increase in CD8$^+$ T cells which can respond to the second infection relative to the first is responsible for the decrease in recovery time for a second infection relative to the first. An additional effect which decreases the recovery time of both the first and second infection is the increased killing rate of effector CTL ($\kappa_{E12}$ and $\kappa_{\hat{E}12}$ are directly proportional to $a_1$).

The total number of CD8$^+$ T cells after the second infection increases superlinearly with avidity (Fig.~\ref{fig:change_avidity}B).  This is because the expansion ratios for both infections increase sublinearly (Fig.~\ref{fig:change_avidity}C), due to two opposing effects on the stimulation rate: a decrease in the number of infected cells required for half-maximal stimulation ($k_{C12}$ and $k_{\hat{C}12}$), and a decrease in the recovery time which decreases the number of infected cells available to stimulate CD8$^+$ T cells.  The latter causes the expansion ratio to once again be lower for the second infection than the first.

\subsubsection{Factor 3: The epitope abundance per infected cell}\label{sec:epitope_abundance}

\begin{figure}[t]
\centering
\includegraphics[width = \textwidth]{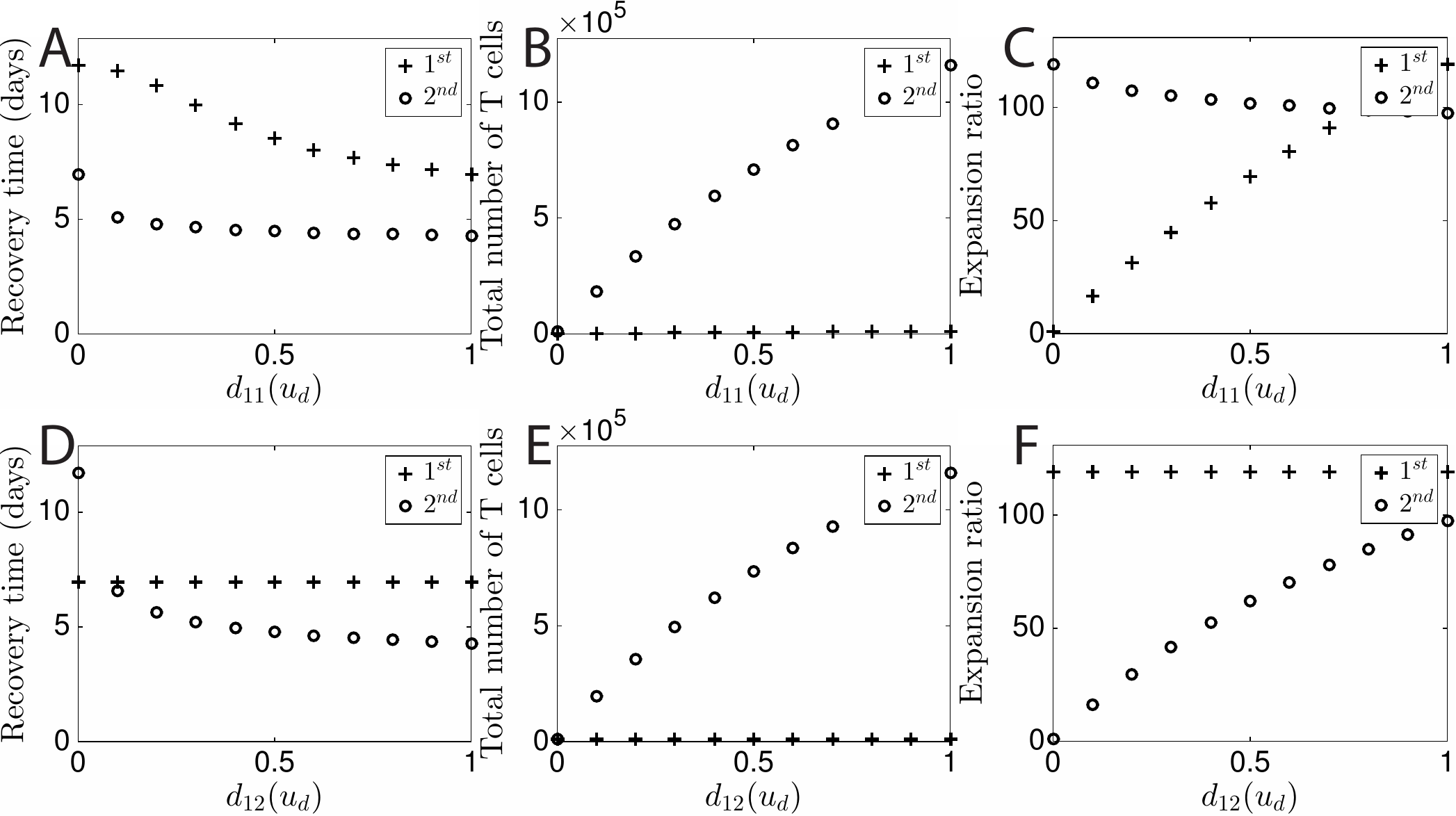}
\caption{The effect of changing the epitope abundance $d_{1q}$ on (A,D) the recovery time, (B,E) the total number of CD8$^+$ T cells 100 days after infection and (C,F) the expansion ratio, for the first and second infection.}
\label{fig:change_epitope_abundance}
\end{figure}

Increasing the epitope abundance per infected cell $d_{1q}$ equally for cells infected with both viruses (i.e. increasing both $d_{11}$ and $d_{12}$) has the same effect as increasing the avidity $a_1$.  The difference between increasing the epitope abundance and increasing the avidity is that we have assumed that the avidity of the pMHC-T cell interaction remains the same regardless of strain, whereas the epitope abundance per infected cell can be unequal for cells infected with different viruses.  This is most obviously true when an epitope is not shared between the two viruses, but even when two strains share an epitope, epitope abundance per infected cell may differ between strains~\citep{Crowe2003}.

We model the situation where the epitope abundance per infected cell $d_{1q}$ differs between strains, and we keep both the avidity of the pMHC-T cell interaction and the initial number of naive CD8$^+$ T cells constant, i.e. $a_{1}  = a_{2} = 1$ and $C_1(0) = \tilde{C}(0)$ (Fig.\ref{fig:cartoon}D).  In the first instance, we vary the epitope abundance per cell infected with strain 1 ($d_{11}$) while keeping the epitope abundance per cell infected with strain 2 ($d_{12}$) constant; then, we vary $d_{12}$ while keeping $d_{11}$ constant, which can be thought of as reversing the order in which the host is infected.

When the epitope abundance per cell infected with strain 1 ($d_{11}$) increases, the recovery time for both the first and second infections decrease (Fig.~\ref{fig:change_epitope_abundance}A), in a similar manner to Fig.~\ref{fig:change_precursor_frequency}A.  Because all parameters for the second virus are held constant, the only difference for the second infection is the increase in CD8$^+$ T cells available for proliferation and differentiation due to the first infection (Fig.~\ref{fig:change_epitope_abundance}C).

The total number of CD8$^+$ T cells increases sublinearly with the epitope abundance per cell infected with strain 1 ($d_{11}$) (Fig.~\ref{fig:change_epitope_abundance}B), as a result of the changes in expansion ratios (Fig.~\ref{fig:change_epitope_abundance}C).  The expansion ratio for the first infection increases sublinearly as the epitope abundance $d_{11}$ increases, because changing $d_{11}$ has the same effect on the number of cells required for half-maximal stimulation of CD8$^+$ T cells ($k_{C11}$) as changing the avidity $a_1$.  The expansion ratio for the second virus decreases as $d_{11}$ increases, because the recovery time, and hence the number of infected cells, decreases, but $k_{C12}$ and $k_{\hat{C}12}$ are constant.

As the epitope abundance per cell infected with the second virus ($d_{12}$) increases, the recovery time for the first infection stays constant, since nothing has changed for that virus.  However, the recovery time for the second infection decreases (Fig.~\ref{fig:change_epitope_abundance}D) due to increased stimulation of naive and memory CD8$^+$ T cells (increased $k_{C12}$ and $k_{\hat{C}12}$) and an increased killing rate for effector CTL ($\kappa_{E12}$ and $\kappa_{\hat{E}12}$). 

Changing $d_{12}$ instead of $d_{11}$ does not change the total number of CD8$^+$ T cells after two infections (Fig.~\ref{fig:change_epitope_abundance}E vs. Fig.~\ref{fig:change_epitope_abundance}B), but changes the expansion ratio (Fig.~\ref{fig:change_epitope_abundance}F).  As $d_{12}$ is increased, nothing changes for the first infection, while the expansion ratio for the second infection increases sublinearly.

Summarising Sec.~\ref{sec:precursor_frequency}--~\ref{sec:epitope_abundance}, when there is a cross-reactive cellular adaptive immune response, for long inter-exposure intervals, the recovery time for the second infection is always less than that of the first.  Whether we increase the precursor frequency $C_1(0)$, avidity $a_1$ or epitope abundance per infected cell $d_{1q}$, the recovery time for the second infection decreases, and the total number of CD8$^+$ T cells after the second infection increases; however, the change in expansion ratio for each infection depends on the factor changing the immune response.  Furthermore, if the strength of the cellular immune response is increased equally for both strains (such as by increasing the precursor frequency or avidity for a common epitope), the expansion ratio for a subsequent infection is less than for a primary infection.  

The result that the recovery time for the second infection decreases as the strength of the cross-reactive cellular adaptive immune response increases is robust to changes in model parameters, as shown in Fig.~\ref{fig:sensitivity_cross_react} in~\ref{sec:cross_react_sensitivity}.

\section{Discussion}
\label{sec:discussion}

We have constructed a multi-strain model of influenza infection within the host including the innate, humoral and cellular adaptive immune responses.  The model for the cellular adaptive immune response includes variable cross-reactivity between strains and a mechanistic model for differentiation of memory CD8$^+$ T cells.  We have used this model to explain our finding \citep{Laurie2015} that a shortening of the second infection was observed when a long interval (10 days or more) separated exposures to heterosubtypic influenza A strains.  As summarised in Fig.~\ref{fig:summary}, both cross-reactivity of the cellular adaptive immune response and the differentiation of effector CTL into memory CD8$^+$ T cells are required to reproduce this shortening of secondary infection.  Our experimental design of successive exposures with heterologous virus with a short inter-exposure interval enables exploration of the roles of the early innate immune response and the subsequent adaptive immune response, as well as the roles of transient effector CD8$^+$ T cells and longer-living memory CD8$^+$ T cells.  We anticipate that some of the identifiability issues which have arisen in previous modelling studies~\citep{Smith2010a,Miao2011,Boianelli2015a} will be ameliorated with this approach; quantifying cross-reactivity and other parameters by fitting our model to the data is the subject of our work in the immediate future.

We also showed that cross-reactivity in the cellular adaptive immune response delays the second infection when there is a short interval between exposures.  However, in the case of delayed infection, stochasticity should be taken into account when comparing the model to experimental data, as suppression of the second virus can manifest as extinction rather than a delay \citep{Cao2015,Yan2016}, in line with the prevented infections observed by \citet{Laurie2015}.

In our model, the strength of the immune response can be changed by modifying the precursor frequency of CD8$^+$ T cells specific to an epitope, the avidity of the pMHC-T cell interaction and the epitope abundance per infected cell.  When an individual is successively infected with two virus strains which share epitopes recognised by the cellular adaptive immune response, changing the strength of the immune response in these three ways impacts the recovery time for both the first and second infection.

Our model predicts that successive infections with heterologous influenza A strains boost the number of epitope-specific CD8$^+$ T cells, such that cellular adaptive immunity is strengthened and the recovery time for successive infections decreases. This is supported by data from mouse experiments~\citep{Christensen2000}, but experiments in other animal systems are lacking.  In a previous experiment, \citet{Laurie2010} measured the viral load time course of up to three successive infections with heterologous influenza A strains in ferrets; similar experiments including T cell measurements would be of great benefit in testing our predictions experimentally.  

In the case of homologous challenge, the antibodies stimulated by the first exposure are equally effective against virus introduced by either exposure, resulting in simultaneous clearance of the virions generated by the first and second exposures (results not shown).  To model this situation, only a single-strain model is needed; the addition of the second strain manifests as a jump discontinuity in the viral load.

Our model has the flexibility to include multiple shared and non-shared epitopes.  This enables investigation of the emergence of immunodominance, the phenomenon by which CD8$^+$ T cells specific to one or two epitopes dominate the immune response.  We can either model the expansion of CD8$^+$ T cells specific to epitopes which are experimentally determined to be dominant, or generate a large range of epitopes with different CD8$^+$ T cell avidities and abundances. In the latter case, the model would have to be expanded to take into account that there are many different CD8$^+$ T cell clones which can respond to a particular epitope with different avidities, as modelled by \citet{Chao2004}.

Our work could also be extended to model the changes in epitope-specific CD8$^+$ T cell numbers over an individual's lifetime~\citep{Quinn2016}, either due to natural infection or vaccination.  Much ongoing work is devoted to the development of next-generation T cell vaccines which would confer immunity against a wider range of strains than current antibody based vaccines~\citep{Brown2009}.  The model would then need to be adapted to take into account the long-term competition of CD8$^+$ T cell clones following repeated infections over an individual's lifetime.  The number of CD8$^+$ T cells in the host is limited, and the total number of memory CD8$^+$ T cells is governed by homeostasis~\citep{Tanchot1995}, causing memory cells produced by new infections to displace those which are no longer stimulated~\citep{Selin1999}.  However, in our model where such considerations are not relevant, increasing the number of infections increases the number of CD8$^+$ T cells without bound.  We could incorporate a saturation term (such as that used by \citet{Wodarz2000}) into the programmed proliferation stage, or explicitly model homeostatic regulation~\citep{Antia2005}.

Many uncertainties remain about the differences between naive and memory CD8$^+$ T cells in terms of the stimulation threshold required for proliferation and differentiation into effector CTL, and in terms of how the effector CTL originating from naive and memory CD8$^+$ T cells differ in parameters such as division rate, total number of divisions, death rate and functionality.  It is clear that memory CD8$^+$ T cells are present in higher numbers than naive CD8$^+$ T cells after primary infection; they also display lysing ability within hours of re-exposure to antigen, in contrast to naive CD8$^+$ T cells which gradually gain functionality over days~\citep{Barber2003,Badovinac2006}.  However, the classic assumption that memory CD8$^+$ T cells have a lower stimulation threshold than naive CD8$^+$ T cells has been questioned~\citep{Mehlhop-Williams2014,Carpenter2016}.  Conventionally, it is also thought that the division rate of memory CD8$^+$ T cells is faster than that of naive CD8$^+$ T cells~\citep{Veiga-Fernandes2000,Veiga-Fernandes2004,Badovinac2006}, and that a greater proportion of effector CTL originating from memory CD8$^+$ T cells survive the contraction phase than do those originating from naive CD8$^+$ T cells~\citep{Grayson2002}.  However, other studies have found that when naive and memory CD8$^+$ T cells are co-transferred into naive hosts before antigen stimulation, naive CD8$^+$ T cells proliferate more quickly than memory CD8$^+$ T cells~\citep{Martin2012}, and that as CD8$^+$ T cells are repeatedly stimulated, their ability to proliferate decreases~\citep{Wirth2010}.  It was also found that a greater proportion of effector CTL originating from naive CD8$^+$ T cells survive the contraction phase than those originating from memory CD8$^+$ T cells~\citep{Martin2012}.

As a result of this uncertainty it is unclear how the number of infected cells required for half-maximal stimulation of naive and memory CD8$^+$ T cells, $k_C$ and $\tilde{k}_C$, and the lysing rates by naive and memory CD8$^+$ T cells, $\kappa_E$ and $\tilde{\kappa_E}$, should differ.  Given that one of the hypotheses for the conflicting results regarding differences between naive and memory T cells is that the experiments transfer different numbers of naive and memory CD8$^+$ T cells to the hosts (making initial CD8$^+$ T cell numbers a confounding factor~\citep{Martin2012}), modelling could potentially resolve the differences between naive and memory T cell characteristics.  

Furthermore, we have assumed that the lysing rate ($\kappa_E$, $\kappa_{\hat{E}}$) is directly proportional to avidity, while the number of infected cells required for half-maximal stimulation of naive/memory CD8$^+$ T cells ($k_C$, $k_{\hat{C}}$) is inversely proportional to avidity.  Biologically, the stimulation threshold for cytolysis is much lower than that for proliferation and differentiation~\citep{Faroudi2003}; we do not know whether the ratio between the two remains fixed between epitopes, such that they can be described with a single avidity parameter, as assumed by some previous models~\citep{Chao2004}.  Furthermore, because low-avidity CD8$^+$ T cells can still be recruited into the immune response but cease proliferation prematurely \citep{Zehn2009}, we can model the effect of avidity by changing the number of proliferation stages $n_{Ej}$ rather than the number of infected cells required for half-maximal stimulation of naive/memory CD8$^+$ T cells ($k_C$, $k_{\hat{C}}$).  This does not change the conclusion that increasing the avidity decreases the recovery time of the second infection and increases the total number of CD8$^+$ T cells after the second infection (results not shown).

There are other aspects of the cross-reactive cellular adaptive immune response which we have not modelled.  For example, we have not taken into account different types of memory CD8$^+$ T cells and their migration patterns in the host. This could not only influence their activation rate and lysing rate~\citep{Shin2013}, but could also introduce delays in lysing activity due to migration time.  In fact, a recent study by~\citet{Zarnitsyna2016} has shown that a model including two types of CD8$^+$ T cells -- resident T cells and central memory T cells -- can explain the shortening of a second infection with strains inducing a cross-reactive immune response, with resident T cells being more effective on a shorter timescale, and central memory T cells acting on a longer timescale.  The delay introduced by migration in the model by~\citet{Zarnitsyna2016} has very similar effects to the refractory memory T cell state in our model, while resident T cells in the aforementioned model are very similar to effector CD8$^+$ T cells in our model; consequently, our model which does not explicitly include the location of T cells delivers similar qualitative results to the location-dependent model.  However, quantification of parameters of different types of T cells, including migration parameters, may prove important for quantitative prediction of infection outcomes and immune boosting.  Moreover, it is of great interest to identify situations in which location dependence of T cells is essential.

We could also model differing avidity within a pool of CD8$^+$ T cells, and/or extend our model to include CD4$^+$ T helper cells, which have effects ranging from maintenance of T cell function~\citep{Wherry2004} to lysis of infected cells~\citep{Brown2012}.  Because CD4$^+$ cells are stimulated by peptide-MHC class II complexes, they are only activated by professional antigen-presenting cells~\citep{Luckheeram2012}; hence, they are stimulated by different epitopes from CD8$^+$ T cells~\citep{Bui2007}, and may be stimulated to different extents.  Our model could be extended to capture different types of cross-reactivity and their net effect on infection outcomes.

\section*{Acknowledgements}
Ada~ W.~C.~Yan is supported by an Australian Postgraduate Award; Pengxing~Cao is supported by a National Health and Medical Research Council (NHMRC) funded Centre for Research Excellence in Infectious Diseases Modelling to Inform Public Health Policy (1078068); Jane~ M.~Heffernan is supported by the Natural Sciences and Engineering Research Council of Canada and the York Research Chair program; J.~McVernon is supported by an NHMRC Career Development Fellowship (CDF1061321); Nicole~L.~LaGruta is a Sylvia and Charles Viertel Senior Medical Research Fellow and is funded by an NHMRC Program grant (AI1071916); James~ M.~McCaw is supported by an Australian Research Council Future Fellowship (FT110100250). The Melbourne WHO Collaborating Centre for Reference and Research on Influenza is supported by the Australian Government Department of Health.

\begin{landscape}
\begin{center}
\begin{table}
\begin{center}
  \begin{tabular}{ lp{9cm}ll }
\hline\noalign{\smallskip}
  Parameter & Description & Value & Units \\  
\noalign{\smallskip}\hline\noalign{\smallskip}
      $\beta$ & infection rate of target cells by virions & $5\times10^{-7}$~\citep{Cao2016}& virion$^{-1}$ day$^{-1}$\\ 
  $g$ & target cell regrowth rate & 0.8~\citep{Cao2016} & day$^{-1}$\\ 
  
    $p_V$ & viral production rate & 12.6~\citep{Cao2016}& virion infected cell$^{-1}$ day$^{-1}$   \\ 
    $p_F$ & interferon production rate & $1 \times 10^{-5}$~\citep{Cao2016}& $u_F$ infected cell$^{-1}$ day$^{-1}$ \\

    $\delta_I$ & infected cell decay rate & 2~\citep{Bocharov1994}& day$^{-1}$\\ 
    $\delta_V$ & virion decay rate & 5~\citep{Cao2016} & day$^{-1}$ \\ 
    $\delta_F$ & interferon decay rate & 2~\citep{Pawelek2012} & day$^{-1}$\\  

    $\kappa_F$ & killing rate of infected cells by natural killer cells & 2.5~\citep{Cao2016} & $u_F^{-1}$ day$^{-1}$ \\    

\noalign{\smallskip}\hline
  \end{tabular}
\end{center}
\caption{Parameter values relating to infection and the innate immune response.  $u_F$ is an arbitrary unit for the amount of interferon.}
\label{table:parameters}
\end{table}
\end{center}

\begin{center}
\begin{table}
\begin{center}
  \begin{tabular}{ lp{10cm}p{5cm}l }
\hline\noalign{\smallskip}
  Parameter & Description & Value & Units \\  
\noalign{\smallskip}\hline\noalign{\smallskip}
    $p_A$ & antibody production rate & $0.8$ & $u_B^{-1}$ day$^{-1}$ \\
        $\delta_A$ & antibody decay rate & 0.04~\citep{Bocharov1994,Lee2009a}& day$^{-1}$ \\
    $\delta_B$ & plasmablast and plasma cell decay rate & 0.1~\citep{Bortnick2013}& day$^{-1}$ \\
        $\kappa_A$ & neutralisation rate of virions by antibodies & 3~\citep{Miao2010}& (pg/mL)$^{-1}$ day$^{-1}$ \\
        $k_B$ & number of virions for half-maximal stimulation of naive B cells & $2\times 10^5$ & virion\\
        $\beta_B$ & maximal stimulation rate of naive B cells & 1 & day$^{-1}$\\
    $\tau_B$ & total proliferation time of plasmablasts& 3~\citep{Marchuk1991,Sze2000}& day \\
        $n_B$ & number of plasmablast division cycles & 5~\citep{Marchuk1991,Sze2000}& division \\    
    \noalign{\smallskip}\hline
  \end{tabular}
\end{center}
\caption{Parameter values relating to the humoral adaptive immune response.  $u_B$ is an arbitrary unit for the number of B cells.}
\label{table:parameters_abs}
\end{table}
\end{center}

\begin{center}
\begin{table}
\begin{center}
  \begin{tabular}{ lp{10cm}p{5cm}l }
\hline\noalign{\smallskip}
  Parameter & Description & Value & Units\\  
\noalign{\smallskip}\hline\noalign{\smallskip}
        $\tilde{k}_{C}$ & number of infected cells required for half-maximal stimulation of naive/memory CD8$^+$ T cells (baseline value) & $5 \times 10^6$ &infected cells $u_a u_d$ \\
                $\tilde{\kappa}_{E}$ & lysing rate of infected cells by effector CTL (baseline value)& $3 \times 10^{-5}$ & effector cell$^{-1}$ day$^{-1} u_a^{-1}u_d^{-1}$ \\
                        $\beta_C$,$\beta_{\hat{C}}$ & maximal stimulation rate of naive/memory CD8$^+$ T cells & 1 & day$^{-1}$\\
        $\delta_E$,$\delta_{\hat{E}}$ & decay rate of effector CTL & 0.6~\citep{Veiga-Fernandes2000}& day$^{-1}$ \\
        $\tau_M$ & mean refractory time for memory CD8$^+$ T cells & 14~\citep{Kaech2002}& day \\
    $\tau_E$,$\tau_{\hat{E}}$ & total proliferation time of effector CTL& 6~\citep{Lehmann-Grube1988}& day \\
    $n_E$,$n_{\hat{E}}$ & number of effector CTL division cycles &20~\citep{vanStipdonk2001}& division \\
        $\epsilon$ & proportion of effector CTL which become (refractory) memory CD8$^+$ T cells& 0.02 ~\citep{Murali-Krishna1998,DeBoer2001,Chao2004}\\
        \noalign{\smallskip}\hline
  \end{tabular}
\end{center}
\caption{Parameter values relating to the cellular adaptive immune response.  $u_a$ is an arbitrary unit for avidity; $u_d$ is an arbitrary unit for epitope abundance.}
\label{table:parameters_cd8}
\end{table}
\end{center}

\begin{center}
\begin{table}
\begin{center}
  \begin{tabular}{ llll }
\hline\noalign{\smallskip}
  Parameter & Description & Value & Units\\  
  $V(0)$ & virion & 10 & virion\\ 
  $T(0)$ & target cell & $7\times10^7$~\citep{Cao2015,Petrie2013}& target cell \\ 
  $B_{0}(0)$ & naive B cell & 10 & $u_{B}$ \\
    $\tilde{C}(0)$ & naive CD8$^+$ T cell & 100~\citep{Blattman2002}& CD8$^+$ T cell \\
          \noalign{\smallskip}\hline
  \end{tabular}
\end{center}
\caption{Initial values.   All unlisted initial values are zero. $u_B$ is an arbitrary unit for the number of B cells.}
\label{table:iv}
\end{table}
\end{center}
\end{landscape}

\newpage
\appendix

\section{Sensitivity analysis}

We wish to see whether the qualitative relationships between the recovery time for a second infection and the degree of cross-reactivity in the cellular adaptive immune response as presented in the main text still hold when model parameters are changed.  To investigate this, we vary chosen parameters one at a time, and reproduce selected figures from the main text for different values of these parameters.  The parameters varied are 

\begin{enumerate}
\item $p_V$, the production rate of virions from cells infected with either virus; 
\item $p_F$, the production rate of interferon from cells infected with either virus; and 
\item $\kappa_A$, the neutralising rate of virions of either strain by antibodies.
\end{enumerate}  

These parameters vary three major features of the system: viral replication in the absence of the immune response, the innate immune response, and the humoral adaptive immune response.  The system features can be varied using other choices of parameters, although some of these are equivalent in terms of their effect on the viral load. For example, changing $\kappa_F$ while keeping $p_F$ constant is equivalent to changing $p_F$ while keeping $\kappa_F$ constant; changing $p_A$ while keeping $\kappa_A$ constant is equivalent to changing $\kappa_A$ while keeping $p_A$ constant. 

The parameters $p_V$, $p_F$ and $\kappa_A$ are varied over the ranges $[5.0,101]$ virion infected cell$^{-1}$ day$^{-1}$, $[10^{-7},10^{-4}]$ $u_F$ infected cell$^{-1}$ day$^{-1}$ and $[3 \times 10^{-1},3 \times 10^7]$ (pg/mL)$^{-1}$ day$^{-1}$ respectively.  The lower bound for $p_V$ corresponds to $R_0 \approx 2$, which is the lower limit of experimentally obtained values of $R_0$ for within-host influenza infection \citep{Smith2010}.  Above the upper bound for $p_V$, the number of target cells drops below one for a single infection, indicating that the probability of target cells becoming extinct would be high if stochasticity of target cell dynamics were to be taken into account.  Because target cell regrowth is proportional to the number of target cells in the model, if the number of target cells drops to zero, regrowth does not occur, and a second infection is not possible.  This indicates a likely breakdown in validity of the model in this region of parameter space.  To increase $p_V$ while avoiding target cell depletion, we could simultaneously increase $p_F$, the production rate of interferon from cells infected with either virus; however, for the purposes of this single-parameter sensitivity analysis, we exclude this region of parameter space.

Below the lower bound for $p_F$, the number of target cells drops below one for a single infection, such that a second infection is not possible.  (To decrease $p_F$ while avoiding target cell depletion, we could simultaneously decrease $p_V$.)  Above the upper bound for $p_F$, the innate immune response decreases the infected cell and viral load equilibrium to the point where effective cellular and humoral adaptive immune responses are not induced, resulting in chronic infection.  As we are interested in the dynamics of acute infection, we consider this region of parameter space biologically implausible and exclude it from our analysis.

Below the lower bound for $\kappa_A$, the humoral adaptive immune response is not strong enough to resolve the infection in the absence of the cellular adaptive immune response, which is at odds with experimental results \citep{Kris1988,Wells1981,Yap1979}.

\subsection{For a long inter-exposure interval, cross-reactivity and memory shorten the recovery time for the second infection}\label{sec:duration_sensitivity}

\begin{figure}[t]
\centering
\includegraphics[width = \textwidth]{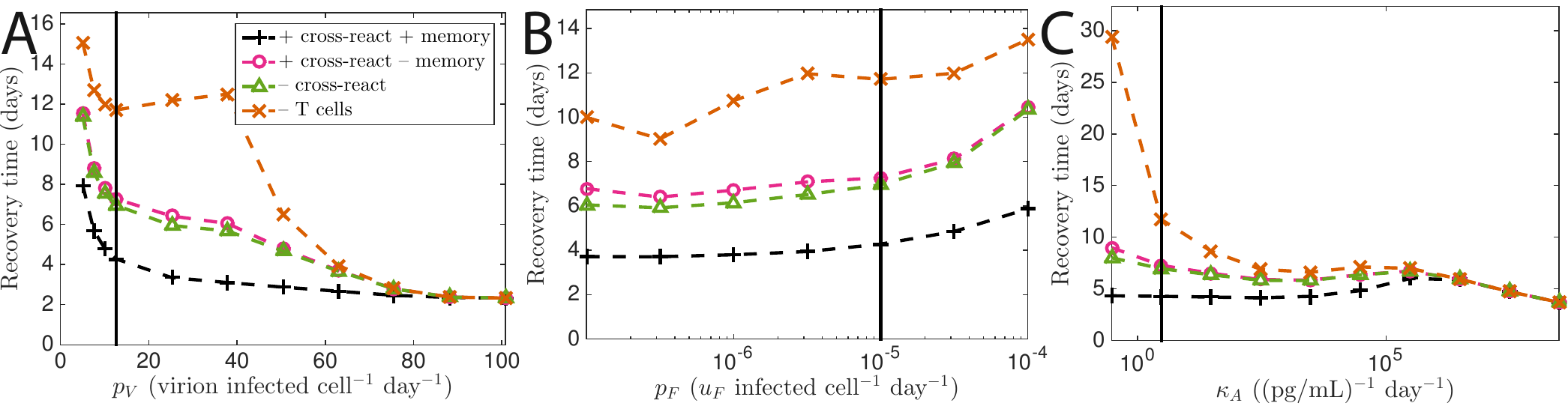}
\caption{Recovery time for the second infection for a 100-day inter-exposure interval as the parameter on the $x$-axis is varied, for four cases: (1) cross-reactivity between the two strains and memory CD8$^+$ T cells present (black line with crosses); (2) cross-reactivity but no memory CD8$^+$ T cells (magenta line with circles); (3) no cross-reactivity between the strains (green line with triangles); (4) no CD8$^+$ T cells (orange line with crosses).  The vertical line indicates the value used for Fig.~\ref{fig:stats}C of the main text.  For a 100-day inter-exposure interval, the recovery time for a single infection is the same as that for the case without cross-reactivity (green line with triangles).}
\label{fig:sensitivity_recovery_time_iei}
\end{figure}

In the main text, we observed that for a long inter-exposure interval (10--14 days), cross-reactivity and memory are required to shorten the recovery time of a second infection relative to that of a first infection (Fig.~\ref{fig:stats}C).  For the purposes of the sensitivity analysis, we will keep the inter-exposure interval constant at 100 days as $p_V$, $p_F$ and $\kappa_A$ are varied.  

For all values of $p_V$ (Fig. \ref{fig:sensitivity_recovery_time_iei}A), $p_F$ (Fig. \ref{fig:sensitivity_recovery_time_iei}B) and $\kappa_A$ (Fig. \ref{fig:sensitivity_recovery_time_iei}C), both cross-reactivity and memory are required to shorten the recovery time of the second infection relative to the first infection, which indicates that our primary findings are robust to uncertainty.  However, this effect is smaller in some regions of parameter space, such as when $p_V$, the production rate of virions from infected cells, is high.  In this region of parameter space, the recovery time is hardly affected if the cellular adaptive immune response is removed (orange line with crosses); this is because the number of infected cells becomes so large for large values of $R_0$ that target cell depletion becomes the dominant factor in resolving infection.  Similarly, for high values of $\kappa_A$, the recovery time for the cases with cross-reactivity converge to that for the case without cross-reactivity because the humoral adaptive immune response efficiently resolves the infection, such that the role of the cellular adaptive immune response is small.  This is evidenced by the little change in recovery time when the cellular adaptive immune response is removed altogether.  However, in the regions of parameter space where the cellular adaptive immune response does have a significant effect in resolving infection, cross-reactivity decreases the recovery time, leaving our conclusions intact.

\subsection{For a long inter-exposure interval, cross-reactivity and memory are required to boost the total number of CD8\texorpdfstring{$^+$}{+} T cells after the second infection}\label{sec:total_T_sensitivity}

\begin{figure}[t]
\centering
\includegraphics[width = \textwidth]{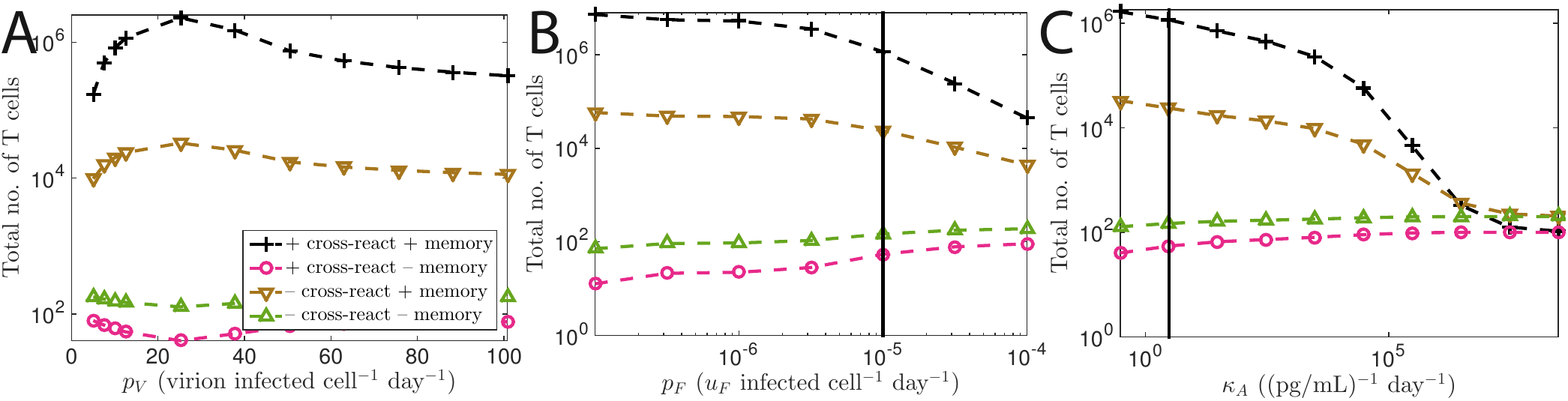}
\caption{Total number of CD8$^+$ T cells 100 days after the second infection, for a 100-day inter-exposure interval as the parameter on the $x$-axis is varied, for four cases: (1) cross-reactivity between the two strains and memory CD8$^+$ T cells present (black line with crosses); (2) cross-reactivity but no memory CD8$^+$ T cells (magenta line with circles); (3) no cross-reactivity but memory CD8$^+$ T cells present (brown line with triangles); (4) no cross-reactivity or memory CD8$^+$ T cells (black line with triangles). The vertical line indicates the value used for Fig.~\ref{fig:T}A of the main text.  The total number of CD8$^+$ T cells is defined as $C_1 + E_1 + M_1 + \hat{C}_1 + \hat{E}_1$.}
\label{fig:sensitivity_total_T}
\end{figure}

In the main text, we observed that for a long inter-exposure interval (10--14 days), both cross-reactivity and memory are required to boost the total number of CD8$^+$ T cells after the second infection (Fig.~\ref{fig:T}A).

For all values of $p_V$ (Fig.~\ref{fig:sensitivity_total_T}A) and $p_F$ (Fig.~\ref{fig:sensitivity_total_T}B), the total number of CD8$^+$ T cells after the second infection is boosted when both cross-reactivity and memory are present, which indicates that our primary findings are robust to uncertainty in these two parameters.   

For low values of $\kappa_A$ (Fig.~\ref{fig:sensitivity_total_T}C), the total number of CD8$^+$ T cells after the second infection is greater when both cross-reactivity and memory CD8$^+$ T cells are present.  However, for high values of $\kappa_A$, the total number of CD8$^+$ T cells after the second infection is marginally greater when the two strains are not cross-reactive.  We may understand this as follows.  We have chosen to keep the initial number of naive CD8$^+$ T cells per pool the same (100 cells) in both situations, but in the cross-reactive case, there is one pool of CD8$^+$ T cells whereas in the non-cross-reactive case, there are two pools of CD8$^+$ T cells.  Hence, the total initial number of CD8$^+$ T cells is greater in the non-cross-reactive case.  For low values of $\kappa_A$, the increased expansion of CD8$^+$ T cells in the cross-reactive case compensates for this initial difference.  However, when $\kappa_A$ is high, the humoral adaptive immune response efficiently clears the infection, such that only a small number of CD8$^+$ T cells is stimulated to proliferate and differentiate.  Hence, the initial difference in the total number of CD8$^+$ T cells dominates, and the total number of CD8$^+$ T cells 100 days after the second infection is less in the cross-reactive case.  However, in this region of parameter space, the expansion of CD8$^+$ T cells in response to infection, and their effect on the viral load, is unrealistically limited.  In contrast, in the region of parameter space where the cellular adaptive immune response has a significant effect in resolving infection, our conclusions are robust to parameter uncertainty.

We note that as $p_V$ (the production rate of virions from infected cells) increases, the total number of CD8$^+$ T cells for the second infection increases, then decreases (Fig.~\ref{fig:sensitivity_total_T}A).  This is because as $p_V$ increases, the number of infected cells at equilibrium increases, but the recovery time decreases; the play-off of the two effects causes the areas under the two curves $\frac{\sum_{q=1}^Q I_q/k_{Cjq}}{1+\sum_{q=1}^QI_q/k_{Cjq}}$ and $\frac{\sum_{q=1}^Q I_q/k_{\hat{C}jq}}{1+\sum_{q=1}^QI_q/k_{\hat{C}jq}}$ to first increase, then decrease.  The areas under the curves are directly proportional to the number of naive/memory T cells which proliferate and differentiate during the course of the second infection (following from Eq.~\ref{eq:dC0dt} and Eq.~\ref{eq:dC0hatdt}).  This results in the total number of CD8$^+$ T cells for the second infection increasing, then decreasing as $p_V$ increases.

Similarly, as $p_F$ (the production rate of interferon) increases, the total number of CD8$^+$ T cells for the second infection decreases (Fig.~\ref{fig:sensitivity_total_T}B) due to a decrease in the number of infected cells at equilibrium which is not completely offset by a simultaneous increase in recovery time.

\subsection{The effect of changing the strength of the cross-reactive immune response on the recovery time for the second infection}\label{sec:cross_react_sensitivity}

In the main text, we described three factors which determine the strength of the cellular adaptive immune response: the precursor CD8$^+$ T cell frequency, the avidity of the pMHC-T cell interaction, and the epitope abundance on infected cells.  We then showed that increasing the strength of the cellular adaptive immune response by increasing the parameters corresponding to these three factors ($C_1(0)$, $a_1$, $d_{11}$ and $d_{12}$) decreases the recovery time for the second infection (Sec.~\ref{sec:cross}).

\begin{figure}[htbp]
\centering
\includegraphics[width = \textwidth]{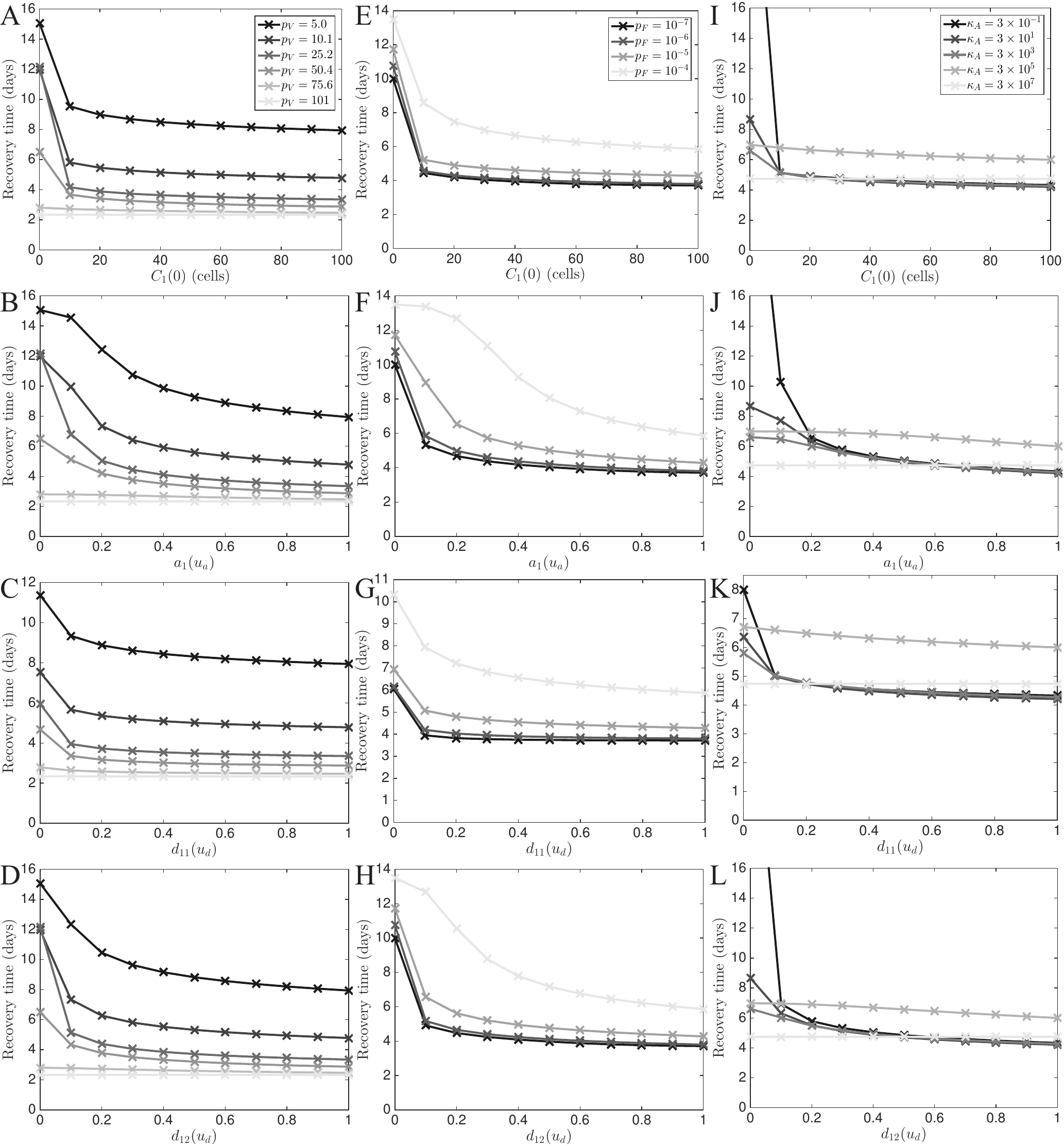}
\caption{Recovery time for the second virus for a 100-day inter-exposure interval, as (from top) the precursor frequency $C_1(0)$, the avidity $a_1$, the epitope abundance on cells infected with the first virus $d_{11}$ and the epitope abundance on cells infected with the second virus $d_{12}$ are varied along the $x$-axes.  Different lines correspond to different values of (left) the production rate of virions from infected cells $p_V$, (centre) the production rate of interferon from infected cells $p_F$ and (right) the neutralisation rate of virions by antibodies $\kappa_A$, as indicated by the legend. $p_V$, $p_F$ and $\kappa_A$ are given in units of virion infected cell$^{-1}$ day$^{-1}$, $u_F$ infected cell$^{-1}$ day$^{-1}$ and (pg/mL)$^{-1}$ day$^{-1}$ respectively.}
\label{fig:sensitivity_cross_react}
\end{figure}

The conclusion that the recovery time decreases as the strength of the cellular adaptive immune response is increased is robust to changes in the production rate of virions ($p_V$) (Figs.~\ref{fig:sensitivity_cross_react}A-D), although the reduction in recovery time decreases, and finally becomes negligible, for large values of $p_V$.  This is because for large values of $p_V$, resolution of the infection is driven by target cell depletion rather than the cellular adaptive immune response.  Nevertheless, our conclusions hold in the region of parameter space where the cellular adaptive immune response has a significant effect on the resolution of the infection.

The conclusion that the recovery time decreases as the strength of the cellular adaptive immune response is increased is robust to changes in the production rate of interferon from infected cells ($p_F$) (Figs.~\ref{fig:sensitivity_cross_react}E-H).  The same conclusion is also robust to changes in the neutralisation rate of virus by antibodies ($\kappa_A$) (Figs.~\ref{fig:sensitivity_cross_react}I-L), although for a fixed cellular adaptive immune response parameter value (e.g. $C_1(0)$), the relationship between $\kappa_A$ and the recovery time is not monotonic; the shapes of the curves also change as $\kappa_A$ increases.  

In detail, in Figures~\ref{fig:sensitivity_cross_react}I and K, we see that for a constant non-zero value of $C_1(0)$ and $d_{11}$, for low values of $\kappa_A$, the recovery time does not change much as $\kappa_A$ is changed; for $\kappa_A = 3 \times 10^5$ (pg/mL)$^{-1}$ day$^{-1}$, the recovery time jumps to a much higher value; and for $\kappa_A = 3 \times 10^7$ (pg/mL)$^{-1}$ day$^{-1}$, the recovery time decreases again.  This is because for high values of $\kappa_A$, the humoral adaptive immune response dominates the resolution of the infection, preventing sufficient stimulation of the cellular adaptive immune response.  Consequently, for high numbers of precursor CD8$^+$ T cells, increasing $\kappa_A$ can actually increase the recovery time of the second infection, which is consistent with the `bump' shown in Fig.~\ref{fig:sensitivity_recovery_time_iei}C.  The ordering of the lines in Figs.~\ref{fig:sensitivity_cross_react}I and K is similar because the parameters $C_1(0)$ (the precursor CD8$^+$ T cell frequency) and $d_{11}$ (the epitope abundance on cells infected with the first virus) both only change the initial conditions for the second virus rather than rate parameters; the ordering is different for the other two parameters $a_1$ and $d_{12}$, but the cause of the non-monotonic ordering is the same.  Regardless, for each line representing an individual value of $\kappa_A$, increasing each of the cellular adaptive immune response parameters $C_1(0)$, $a_1$, $d_{11}$ and $d_{12}$ decreases the recovery time, which is consistent with the findings in the main text.

\section*{References}
\bibliography{library_CD8} 

\end{document}